\documentstyle[12pt,psfig,subfigure,epsf]{article}

\newcommand{\rf}[1]{(\ref{#1})}
\newcommand{\bea}{\begin{eqnarray}}
\newcommand{\eea}{\end{eqnarray}}

\newcommand{\e}{{\mathrm{e}}}
\renewcommand{\d}{{\mathrm{d}}}

\newcommand{\g}{\gamma}

\renewcommand{\a}{\alpha}

\newcommand{\m}{\mu}

\newcommand{\ep}{\varepsilon}

\newcommand{\del}{\delta}

\newcommand{\oh}{\frac{1}{2}}
\newcommand{\oq}{\frac{1}{4}}

\newcommand{\ra}{\right\rangle}
\newcommand{\la}{\left\langle}
\newcommand{\prt}{\partial}

\newcommand{\cD}{{\cal D}}
\newcommand{\cS}{{\cal S}}

\newcommand{\cT}{{\cal T}}

\newcommand{\vev}[1]{\langle #1 \rangle}

\newcommand{\hstackrel}[2]{ 
    \stackrel{\hbox{\scriptsize #1}}{\hbox{\scriptsize #2}} }

\def\void{}
\def\labelmark{}

\newenvironment{formula}[1]{\def\labelname{#1}
\ifx\void\labelname\def\junk{\begin{displaymath}}
\else\def\junk{\begin{equation}\label{\labelname}}\fi\junk}%
{\ifx\void\labelname\def\junk{\end{displaymath}}
\else\def\junk{\end{equation}}\fi\junk\labelmark\def\labelname{}}

{\ifx\void\labelname\def\junk{\end{array}\end{displaymath}}
\else\def\junk{\end{array}\right.\end{equation}}
\fi\junk\labelmark\def\labelname{}\def\junk{}
}

\newcommand{\beq}{\begin{formula}}
\newcommand{\eeq}{\end{formula}}
\newcommand{\beqv}{\begin{formula}{}}

\begin{document}
\topmargin 0pt
\oddsidemargin 5mm
\headheight 0pt
\headsep 0pt
\topskip 9mm

\hfill    EPHOU-97-006

\hfill    NBI-HE-97-19

\hfill    TIT/HEP--371

\hfill June 1997

\begin{center}
\vspace{24pt}
\textbf{\large The quantum space-time of $c=-2$ gravity}

\vspace{24pt}

\textsl{ J. Ambj\o rn}\footnote{ The Niels Bohr Institute
Blegdamsvej 17, DK-2100 Copenhagen \O , Denmark.},  
\textsl{K. Anagnostopoulos}$^1$, 
\textsl{T. Ichihara}\footnote{Department of Physics, Tokyo Institute of 
Technology, O-okayama, Meguro, Tokyo, Japan. }, 
\textsl{L. Jensen}$^1$, \\
\textsl{N. Kawamoto}\footnote{Department of Physics, Hokkaido University,
Sapporo, Japan.}, 
\textsl{Y. Watabiki}$^2$ and \textsl{K. Yotsuji}$^3$

\end{center}
\vspace{24pt}

%\addtolength{\baselineskip}{0.20\baselineskip}
\vfill

\begin{center}
\textbf{Abstract}
\end{center}

\vspace{12pt}

\noindent
We study the fractal structure of space-time of two-dimensional 
quantum gravity coupled to $c=-2$ conformal matter by means of 
computer simulations. We find that the intrinsic Hausdorff dimension 
$d_H = 3.58 \pm 0.04$. This result supports the conjecture 
$d_H = -2 \alpha_1/\alpha_{-1}$, where $\alpha_n$ is the gravitational 
dressing exponent of a spinless primary field of conformal weight 
$(n+1,n+1)$, and it disfavours the alternative prediction 
$d_H = 2/|\gamma|$. 
On the other hand $\langle l^n \rangle \sim r^{2n}$ 
for $n>1$ with good accuracy, i.e. the the boundary length $l$ 
has an anomalous dimension relative to the area of the surface.

\vfill

\newpage

\section{Introduction}
\label{s:1}

We still do not understand the theory of quantum gravity. In four
dimensions it has been difficult to reconcile quantum theory and
gravity. We have been more successful in two dimensions.  Matrix
models \cite{david,adf,bkkm} and Liouville field \cite{kpz,ddk} theory
allow us to understand a lot about the interplay between
two-dimensional quantum gravity and conformal field theory.

It is sometimes said the two-dimensional quantum gravity will tell us
little about four-dimensional quantum gravity since there are no
gravitons in two dimensions. However, many of the conceptional
problems of a theory of quantum gravity remain the same in two and
four dimensions.  In a certain way two-dimensional quantum gravity is
as ``quantum like'' as a theory can be: In the path integral we
perform a summation of all geometries with weight one\footnote{Except
for a cosmological constant term, which has no dependence on the
derivatives of the metric.}.  This means that there is no classical
background space-time around which we expand and our understanding of
space-time in such a quantum world could indeed contain very important
messages of use in higher dimensions.

From this point of view it is somewhat annoying that precisely the
structure of space-time is the least understood in two-dimensional
quantum gravity. The recent introduction of the so-called transfer
matrix \cite{transfer} allows us to analyse in a satisfactory way the
fractal structure of space-time for pure two-dimensional quantum
gravity \cite{transfer,aw,kawai} and it highlights the fact that the
dimension of space-time in quantum gravity is a {\em dynamical}
quantity. Even if our underlying theory is two-dimensional we cannot be
sure that this is the case for the quantum average. In fact, the
fractal dimension of pure two-dimensional quantum gravity is four !

While the transfer matrix technique works perfectly for pure
two-dimensional quantum gravity it is difficult to implement in the
case of where matter fields are coupled to two-dimensional
gravity. The reason is simple: when performing the sum over
intermediate states, we should not only sum over all geometries of a
certain kind but also over intermediate states of the matter
fields. This has until now been impossible and another strategy has
been followed: the concept of intermediate states is redefined in such
a way that the summation {\em can} be performed \cite{deformation}.
The price paid is that the concept of intermediate state looses its
direct link to the geometry present at the underlying manifold.  To be
more specific, let us consider two-dimensional quantum gravity coupled
to a $c=1/2$ conformal field theory. At a constructive level, this
theory is realized as the theory of dynamical triangulations with
Ising spins placed at the centers of the triangles and the coupling of
the Ising spins at a certain critical value. The geometry is now well
defined via the dynamical triangulations and we can discuss the
propagation of a one-dimensional universe with length $\ell_1$ to
another one-dimensional universe with length $\ell_2$, the two
separated a geodesic distance $D$. In the case of pure gravity this
problem is solved by the transfer matrix method which reduces the
calculation to a summation over successive amplitudes between
universes where the one-dimensional boundaries are separated an
infinitesimal distance. The only technical problem in the case of pure
gravity is that the summation over all intermediate length $\ell$ of
the boundaries has to be performed. This problem was solved in
\cite{transfer}.  When we have Ising spins at the centers of the
triangles we have in addition to sum over possible spin assignments at
the boundaries of length $\ell$. Presently this cannot be done
analytically. This problem can be avoided by redefining what is meant
by the distance relating two one-dimensional loops by only considering
deformations from loops with all spin aligned to other loops where all
spin are aligned (the prescription can be made precise). However, such
loops might not have a well defined {\em geodesic}
distance. Nevertheless, consistent scaling relations can be defined in
terms of this modified distance. If we {\em assume} that the modified
distance is proportional to the real geodesic distance when the
quantum average is performed, we get the prediction that the fractal
dimension of space-time for two-dimensional quantum gravity coupled to
a conformal field theory of central charge $c$ is 
\beq{*1} 
 d_H = \frac{2}{|\g(c)|}, 
\eeq 
where the string susceptibility is given by
\beq{*2} 
 \g (c) = \frac{c-1 - \sqrt{(25-c)(1-c)}}{12} \, .
\eeq 
In particular, for an $(m,m\!+\!1)$ conformal field theory 
one obtains $d_H = 2m$, 
and for the Ising model, which corresponds to $m=3$, 
one gets $d_H = 6$. 
On the other hand, 
for $c=-2$ (a non-unitary conformal theory) 
we get 
\beq{*3}
 d_H(c=-2) =2 , 
\eeq
i.e.\ the fractal dimension $d_H$ is equal to 
the dimension of the underlying manifold.

An alternative prediction of $d_H$ is obtained
by  use  of the diffusion equation in Liouville theory \cite{ksw}:
\beq{*25}
d_H  =  -2\frac{\a_1}{\a_{-1}} = 
2 \times \frac{\sqrt{25-c}+\sqrt{49-c}}{\sqrt{25-c}+ \sqrt{1-c}}\, .
\eeq
The origin of this equation is to be found 
in the analysis of the diffusion equation in Liouville theory \cite{ksw}
and is based on the observation that for random walks 
on a two-dimensional manifolds of area $A$ we expect 
\beq{*26}
{\rm dim}[ \, \la r^2(t) \ra_A \, ] = {\rm dim}[ \, A^{2/d_H} \, ] \, ,
\eeq
where $r(t)$ is the geodesic distance of  the random walk
at the (fictitious) diffusion time $t$ and the average 
refers to the functional integral over geometries and matter. 
In Liouville theory one can use the De-Witt short distance expansion of 
the heat kernel in terms of geodesic distance to deduce \cite{ksw} that
\beq{*27}
{\rm dim}[ \, \la r^2(t) \ra_A \, ] = 
{\rm dim}[ \, A^{-(\a_{-1}/\a_1)} \, ] .
\eeq 
In eq.\ \rf{*27} $\a_{-n}$ denotes the gravitational dressing of 
an $(n\!+\!1,n\!+\!1)$ primary spinless conformal field, i.e.\
$$
\int\! d^2\xi \sqrt{g}\Phi_{n+1}(g) \to
\int\! d^2\xi \sqrt{\hat{g}} \; e^{\a_{-n} \phi} \Phi_{n+1}(\hat{g}) 
~~~{\rm for}~~~
g_{\mu\nu}(\xi) = e^{\phi(\xi)} \hat{g}_{\mu\nu}(\xi),
$$
where $\hat{g}_{\mu\nu}(\xi)$ is the background metric 
and $\Phi_{n}(g)$ satisfies 
$\Phi_{n}(e^\phi \hat{g}) = e^{-n\phi} \Phi_{n}(\hat{g})$. 
The requirement that 
$e^{\a_{-n}\phi}\Phi_{n+1}(\hat{g})$ is a $(1,1)$ conformal field fixes 
\beq{*28}
\a_n = \frac{2n}{1+\sqrt{(25-c-24n)/(25-c)}}\, .
\eeq
It is an important assumption in this derivation that it is legal to 
commute the asymptotic De-Witt expansion of the heat kernel 
and the functional integral over geometry and matter. 
For $c=0$ one obtains $d_H = 4$, in agreement with the transfer matrix 
prediction, while for $c=1/2$ (corresponding to the critical Ising model
coupled to gravity in the transfer matrix formulation) one obtains 
$d_H = (\sqrt{97}+7)/4 = 4.212\ldots $. For $c=-2$ one obtains 
\beq{*29}
  d_H(c=-2) = \frac{3+\sqrt{17}}{2} = 3.561\ldots \, .
\eeq

Except for pure gravity the two predictions disagree. It is the purpose
of the present work to test if any of the two predictions 
is consistent with numerical simulations. The available analytical 
methods have build-in assumptions. 
No assumptions have to be made by a br\^{u}te force 
numerical simulation of the system. Of course numerical methods
have other problems, most notably that of accuracy. Until now the numerical 
simulations have been concentrated on  systems with $c>0$, mainly 
the Ising spin coupled to gravity ($c=1/2$) and the three-state
Potts model coupled to gravity  ($c=4/5$) \cite{suracuse,ajw,aamt,ak}.
The results have so far not been able to support the prediction 
\rf{*1} (i.e.\ $d_H = 6$ and 
10 for the Ising and the three-state Potts model, respectively).
However, it could be  argued that these dimensions are so large 
that it would be very difficult to observe them in numerical 
simulations with the present size of lattices. It is natural 
to require that one should be able to probe lattice distances $r$ such that  
\beq{*50}
1 \ll r \ll N^{1/d_H},
\eeq
where $N$ is the number of triangles or vertices in the triangulation.
For systems with $N < 10^6$ it is clearly problematic to fulfill
\rf{*50}. It is however possible to measure the critical indices of
Ising and three-state Potts models coupled to quantum gravity with
good precision \cite{indices}.  Since many of these critical indices
come from integrated two-point functions which also measure $d_H$ it
is not easy to understand how the critical properties come out right
if \rf{*50} is never satisfied.  Moreover, it was recently
found that correlation functions defined in terms of geodesic distance
scale consistently with the theoretical critical indices if and only
if $d_H\approx 4$ \cite{suracuse,aamt,ak}.  These are indirect
arguments in disfavour of \rf{*1}, but of course not conclusive since
numerical peculiarities could conspire and still allow us to determine
the critical exponents without \rf{*50} ever being fulfilled. However,
in order to avoid this discussion completely it is convenient to turn
to the $c=-2$ system.  This system is in a way the simplest coupled
gravity-matter system. Within the context of dynamical triangulations
it was the first system which could be explicitly solved \cite{bkkm},
apart from pure gravity itself. Quite recently it has been possible to
construct explicitly the two-point function using the transfer matrix
methods \cite{akw} and thereby generalising the results known for pure
gravity \cite{aw}.

The advantage of the choice $c=-2$ is two-fold.  The prediction from
the transfer matrix formulation is $d_H=2$, while it from Liouville
diffusion is 3.562 (see \rf{*3} and \rf{*29}).  These values for $d_H$
are so small that one has no problem satisfying \rf{*50} for the sizes
of systems available on the computer.  Further, $c=-2$ is special
since one does not have to perform Monte Carlo simulations in order to
generate lattice configurations with the correct weight
\cite{kketal}. As we will review below there exists a recursive and
very fast algorithm which allows us to generate directly independent
triangulations. In this way one can use larger systems and obtain
better statistics. This method was first used in \cite{kketal} to
investigate the fractal properties of space--time for the $c=-2$
system coupled to gravity. It was the first numerical confirmation of
the fractal structure of quantum gravity in two dimensions. In this
work the emphasis was put on the use of very large systems in order to
get an unambiguous identification of continuum observables.  Since
then it has been shown that (a): finite size scaling is by far the
most powerful tool for extracting critical properties in
two-dimensional quantum gravity (i.e.\ the situation is similar to the
one for ordinary statistical systems) and (b): the fractal properties
of space-time have an interpretation as critical indices associated
with two-point correlators, precisely as in ordinary statistical field
theory \cite{aw,ajw}. In this article we will show that (a) and (b)
together with the powerful technique of recursive sampling available
for $c=-2$ coupled to quantum gravity makes it possible to determine
$d_H$ with a precision not known before.

The rest of this paper is organised as follows: 
In section 2 we discuss the model as well as 
the observables and their scaling. 
Section 3 outlines the numerical methods used. 
Section 4 contains a summary of the numerical results obtained, 
while section 5 estimates the finite size shift of geodesic distance 
from the theoretical point of view. 
In section 6 we discuss the results and the implications.

The results  strongly  support  \rf{*29}, i.e.\ it lends support to the 
prediction \rf{*25}. 
However, we also find a surprising new scaling law for the boundaries 
in two-dimensional quantum gravity.
A short report of some of the results discussed in this paper has  
appeared in \cite{us}.

\section{The model}
\label{s:m}

\subsection{The partition function}
\label{ss:pf}

Within the framework of dynamical triangulations the conformal theory
of $c$ Gaussian fields $x^\m$ coupled to quantum gravity is described
by the following partition function
\beq{*59}
Z_N = \sum_{T_N \in \cT_N} \frac{1}{\cS_{T_N}} \int \prod_{i=1}^N dx_i \; 
\exp \oh \Bigl(\sum_{(ij)} (x_i-x_j)^2\Bigr)\; 
\del\Bigl(\sum_i x_i\Bigr),
\eeq
where $\cT_N$ denotes the set of triangulations of fixed topology
(which we always assume is spherical) constructed from $N$ triangles.
${\cS_{T_N}}$ is a symmetry factor.  The $c$ independent Gaussian
variables $x^\m_i$ can be viewed as placed at the center of triangle
$i$. They interact with the Gaussian variables at the neighbouring
triangles and $\sum_{(ij)}$ denotes the sum over all such pairs of
triangles.

The Gaussian integration can be performed and one obtains
(up to a constant of proportionality)
\beq{*60}
Z_N  =  \sum_{T_N} \frac{1}{\cS_{T_N}} \left(\det{}' C_{T_N}\right)^{-c/2}\, ,
\eeq
where $C_{T_N}$ is the so-called adjacency matrix of the closed
$\phi^3$-graph $\phi^3(T_N)$ dual to $T_N$. From graph theory it is
known that $\det'\!C_{T_N}$ is equal to the number of rooted spanning
trees in the graph $\phi^3(T_N)$.  Eq.\ \rf{*60} serves as a
definition of a model when $c$ is not a positive integer, in
particular when $c=-2$.  The string susceptibility can be calculated
in this model and it agrees with the continuum calculation in
Liouville theory for a $c=-2$ theory.

It is seen that Eq.\ \rf{*60} is special if $c=-2$ since in this case
we can use the fact that $\det'\!C_{T_N}$ is the number of spanning
trees of $\phi^3(T_N)$, i.e. the number of possible ways to cut the
$\phi^3(T_N)$ graphs of spherical topology into tree diagrams.  The
triangulations in $\cT_N$ are in one-to-one correspondence with the
$\phi^3(T_N)$ connected planar graphs with $N$ vertices and no
external legs. This can be symbolically written as follows:
\beq{*61}
Z_N = \sum_{T_N} \frac{1}{\cS_{T_N}} \! 
      \sum_{\hstackrel{spanning trees}{in $\phi^3(T_N)$}} 1.
\eeq

\subsection{Trees and rainbows}
\label{ss:tar}

\begin{figure}[htb]
\begin{center}
\subfigure{
\psfig{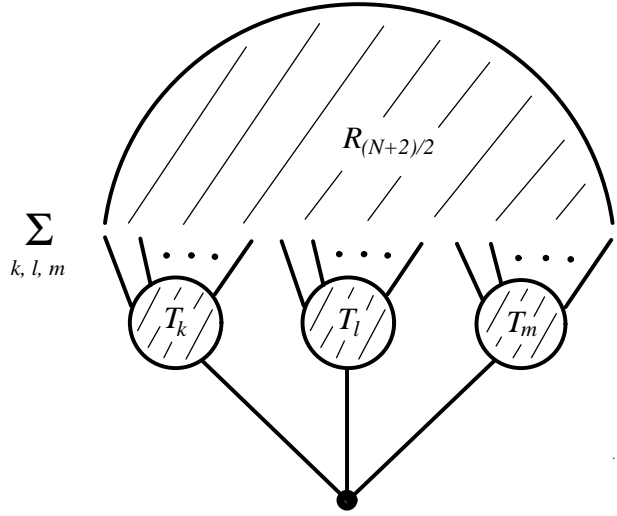} }
\end{center}
\vspace{24pt}
\caption{Partition function described by tree and rainbow diagrams.}
\label{tree0}
\end{figure}

Let us briefly describe the combinatorics associated with the
decomposition of the planar graphs $\phi^3(T_N)$ of spherical topology
into trees and rainbow diagrams.  Let $T_n$ and $R_n$ be the number of
rooted dual tree diagrams with $n+1$ external legs and the number of
rainbow diagrams with $n$ lines, respectively.  Especially, we have
$T_1 = R_0 = 1$.  Here, we mark one of the legs for each tree diagram
and for each rainbow diagram in order to break the symmetry.  Since
any planar closed $\phi^3$ graph can be obtained from a spanning tree
by connecting the external vertices of the tree by rainbow diagrams we
can write \rf{*61} as \cite{bkkm} (See fig.\ \ref{tree0})
\beq{*62}
Z_N  =  
\frac{1}{3N} \! \sum_{ \hstackrel{$k,l,m \ge 1$}{with $k+l+m=N+2$} } \! 
T_{k} T_{l} T_{m} R_{(N+2)/2} 
\, .
\eeq

\begin{figure}[htb]
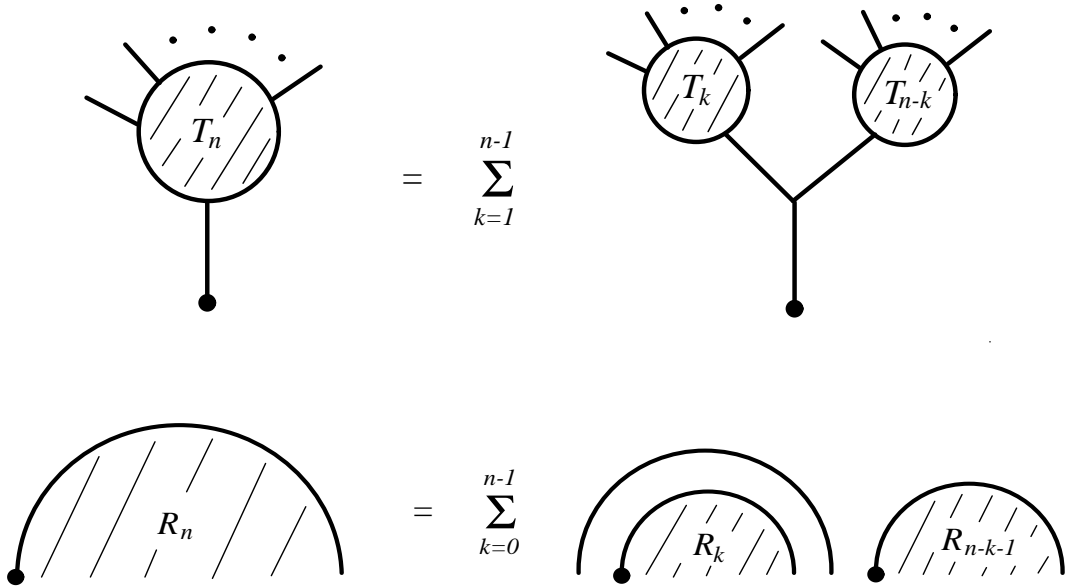

\begin{center}
\subfigure{
\psfig{figure=tree_a.eps,width=12cm,angle=90} }
\end{center}
\begin{center}
\subfigure{
\psfig{figure=tree_b.eps,width=14cm,angle=90} }
\end{center}
\caption{Graphical representation of Schwinger--Dyson equations for 
({\it a}) tree diagrams, ({\it b}) rainbow diagrams.}
\label{tree}
\end{figure}

To calculate $T_n$ and $R_n$ is easily done as follows. 
The tree diagrams satisfy the graphical Schwinger--Dyson equation 
shown in fig.\ \ref{tree}a, 
i.e.\ $T_n$ satisfies 
\beq{xx1}
T_n = \sum_{k=1}^{n-1} T_k T_{n-k} \, .
\eeq
In the same way the rainbow diagrams satisfy the graphical
Schwinger--Dyson equation
of fig.\ \ref{tree}b, leading to an identical equation 
\beq{xx2}
R_n = \sum_{k=0}^{n-1} R_k R_{n-k-1} \, .
\eeq
In order to solve the Eqs.\ \rf{xx1} and \rf{xx2}, 
we introduce the generating functions for the tree diagrams 
and the rainbow ones as 
\beq{*t0}
T(z) = \sum_{n=1}^\infty T_n z^{n-1} ,
~~~~~~
R(z) = \sum_{n=0}^\infty R_n z^{n} .
\eeq
Then, Eqs.\ \rf{xx1} and \rf{xx2} are written 
by using the generating functions as 
\bea
&&T(z) = 1 + z T(z)^2 , ~~~~ ( T(0)=1 ) ,
\label{*t1}\\
&&R(z) = 1 + z R(z)^2 , ~~~~ ( R(0)=1 ) .
\nonumber
\eea
The solutions of \rf{*t1} are 
\beq{*t2}
T(z) = R(z) = \frac{1}{2z} \Bigl( 1-\sqrt{1-4z} \Bigr) . 
\eeq
Therefore, one finds 
\beq{*t3}
T_n = R_{n-1} = \frac{(2n-2)!}{n!(n-1)!} \, . 
\eeq
Using the relation \rf{xx1}, 
the partition function \rf{*62} finally can be written 
as a simpler expression, 
\beq{*63}
Z_N  =  
\frac{1}{N+2} T_{N+1} R_{(N+2)/2} 
%\Bigl( \sum_{ \hstackrel{tree diagrams}{with $N+2$ legs} } 1 \, \Bigr) 
%\Bigl( \sum_{ \hstackrel{rainbow diagrams}{with $(N+2)/2$ lines} } 1 \, \Bigr) 
\, . 
\eeq
%where the first summation is over all rooted $\phi^3$ tree diagrams 
%with $N+2$ external legs 
%and the second summation is over all rainbow diagrams 
%with $(N+2)/2$ lines. 
%$1/(N+2)$ is considered as the symmetry factor which comes from 
%connecting the tree diagram and the rainbow diagram. 

\subsection{Observables}

We define the fractal structure of quantum gravity in the 
following way. Let us fix the space-time volume $V$.
The average volume $S_V(R)$ of a spherical shell  of 
radius $R$ is then 
\beq{*70}
S_V(R) = 
\frac{1}{Z(V)} \int_V \! \cD[g]\cD \phi \,
\e^{-S} \!\int\! d^2 \xi \sqrt{g} \; \del (D_g(\xi,\xi_0)-R),
\eeq
where $Z(V)$ is the partition function of gravity coupled to matter,
with space-time constrained to have volume $V$, $\int_V$ symbolises
that the integration of metrics fulfilling the same constraint,
$\xi_0$ denotes an arbitrary marked point and $D_g(\xi,\xi_0)$ the
geodesic distance from the marked point $\xi_0$ to $\xi$, measured
with respect to the metric $g$. We define the fractal dimension (or
intrinsic Hausdorff dimension) $d_h$ of the space-time by
\beq{*71}
S_V(R)  \sim  R^{d_h-1} %+ O (R^{d_h})
~~~~
\hbox{for \ $R \sim 0$} .
\eeq
It is important to notice that the limit $R \to 0$ is taken 
{\em after} the functional average is performed. Had we taken the 
limit $R \to 0$ before the functional average we would of course have 
obtained the result ``$d_h$''$= 2$ since each manifold is two-dimensional.
However, the limit $R \to 0$ {\it does} not commute with 
the functional integral. It turns out that no matter how small $R$ is
there will always be numerous metrics (i.e.\ a set of metrics of 
non-zero measure with respect to $\cD [g]$) with the property that
geodesic spheres of radius $R$ consist of many connected 
components. For such geometries we cannot necessarily expect 
a growth as slow as $R$. The precise growth of $S_V(R)$ for 
small $R$ becomes a subtle question of entropy of different 
metrics and from this description it is obvious that the 
phrase ``fractal dimension'' is  quite appropriate if $d_h >2$.

In the case of pure two-dimensional gravity it is a remarkable fact 
that one can calculate $S_V(R)$ analytically \cite{aw,ajw}
(it can be expressed in 
terms of certain generalised hypergeometric functions). One finds
\beq{*72}
S_V(R) = R^3 f(R/V^\oq),
\eeq  
where $f(0) > 0$ and $f(x) \sim e^{-x^{4/3}}$ for large $x$.  It is
seen that a dimensionless scaling variable $R/V^{1/4}$ appears. For a
general model such a dimensionless scaling variable, $R/V^{1/d_H}$
will define another intrinsic Hausdorff dimension $d_H$.  From
\rf{*71} and \rf{*72} we deduce that $d_h=d_H=4$ in the case of pure
gravity.  For general model we can write \rf{*72} as
\beq{*73}
S_V(R) = V^{1-1/d_H} F_1(R/V^{1/d_H}),
\eeq
where 
\beq{*74}
F_1(x) \sim x^{d_h-1},~~~~x \ll 1,
\eeq
and $F_1(x)$ goes to zero as $e^{-x^{d_H/(d_H-1)}}$ for $x$ going
to infinity.  Eq.\ \rf{*73} has the form of a typical {\em finite size
scaling relation} and we expect it to be valid not only for pure
gravity, but also for gravity coupled to matter.  Since $S_V(R)$ is
easily measured in numerical simulations we can use Eq.\ \rf{*73} and
\rf{*74} to extract $d_h$ and $d_H$.  Using \rf{*73} and \rf{*74}, one
finds
\beq{*71a}
S_V(R) \sim  V^{1-d_h/d_H} R^{d_h-1} , 
~~~~ R/V^{1/d_H} \ll 1 .
\eeq
If space--time for large $V$ has the same fractal properties at all
scales, one expects 
\beq{SmoothCond}
d_h = d_H . 
\eeq
However, in our numerical simulations we {\it do not assume\/} this
property $d_h = d_H$.  It is one of our purposes to check if
eq.~\rf{SmoothCond} is realized for the $c=-2$ model.
 
Let us briefly describe how the above continuum description translates
to the framework of dynamical triangulations.  To a triangulation
$T_N$ we can unambiguously associate a piecewise linear manifold with
a metric dictated by the length assignment $\ep$ to each link.  From a
practical point of view we use instead a graph-theoretical distance
between vertices, links or triangles.  In the limit of very large
triangulations we expect that the different distances when used in
ensemble averages will be proportional to each other.  To be specific
we will in the following operate with a ``link distance'' and a
``triangle distance''.  The link distance between two vertices is
defined as the shortest link-path between the two vertices, while the
triangle distance between two triangles is defined as the shortest
path along neighbouring triangles between the two triangles.  In this
way the triangle distance becomes the link distance in the dual
$\phi^3$ graph.

In the following we will report on the measurement of quantities
related to the fractal structure of quantum space-time: The total
length $\la l \ra$ and the higher moments $\la l^n \ra$ of spherical
shells of (geodesic) radius $r$, and the distribution function
$\rho(r,l)$ which measures the (average) number of {\em connected}
components of the shells of length $l$ and radius $r$.

More precisely, let us consider the class of triangulations $\cT_N$
which are dual to the connected closed $\phi^3$-graphs.  The number of
triangles (or vertices in the $\phi^3$-graphs), $N$, plays the role of
volume. If $\ep$ denotes the link length of the triangles the relation
to the continuum volume is $V \sim \ep^2 N$, and we want to take a
limit where $V$ is fixed while $\ep \to 0$ and $N$ to infinity. We
consider a spherical ball of radius $r$ and its shell for a given
triangulation $T_N$.  The spherical ball consists of all vertices with
link distance $r' \leq r$ and the spherical shell consists of all
vertices with link distance $r$, where the distance is measured from a
given vertex $v_0$ which is considered as the center of the spherical
ball.  In the same way we can define the spherical shell in terms of
triangle distance. We will use both definitions in the following and
we expect that after taking the statistical average they will be
proportional to each other and that they will not affect the universal
properties of correlation functions \cite{sbb}. The spherical shell in
general consists of a number of connected components if we define a
connected component of the shell of vertices as a maximal set of
vertices in the shell where all vertices can be connected via links in
the shell.  If we take the average over all positions of $v_0$ and all
triangulations $T_N$, we get a distribution $\rho_N(l,r)$ of the
length $l$ (measured in link units) of the connected components of the
spherical shells of radius $r$, i.e.
\beq{*75} 
\la l^n \ra_{r,N}  \equiv  \sum_{l=1}^\infty l^n \rho_N(l,r)\, .  
\eeq
In particular we introduce the special notation
$n_N(r) = \la l \ra_{r,N}$, and since $n_N(r)$ is the discretized version 
of $S_V(R)$ we expect the fractal dimension to be
related to $n_N(r)$ by
\beq{*76} 
n_N(r) \sim r^{d_h-1},~~~~~~~1 \ll r \ll N^{1/d_H}\,.  
\eeq 
According to the general scaling arguments mentioned above
\cite{aw,suracuse,ajw}
we expect the following behaviour for $n_N(r)$: 
\beq{*77} 
n_N (r) \sim N^{1-1/d_H} F_1(x),~~~~~x = \frac{r}{N^{1/d_H}}\, , 
\eeq 
and we expect $F_1(x)$ to behave
as $x^{d_h-1}$ for small $x$ and to fall off rapidly when $x \gg 1$.

\section{Numerical method}
\label{s:nm}
\subsection{The recursive algorithm}
\label{ss:ra}

The recursive algorithm takes advantage of the factorisation property
of the partition function (see Eq.~\rf{*62}) in order to construct a
typical configuration of $c=-2$ gravity. One constructs a rooted
$\phi^3$ tree with the correct probability and then connects the outer
links of the tree with a rainbow diagram also constructed with the
correct probability. 
According to Eq.~\rf{xx1} the branching probability 
to divide a rooted tree diagram with $n+1$ external legs
into two different rooted tree diagrams with $k+1$ and $n-k+1$ (``1''
counts the root) external legs is given by
\beq{*k1}
w(n,k) = \frac{T_k T_{n-k}}{T_n}\, .
\eeq
If we want to construct a surface with $N$ triangles ($N$ must be
even) we need to construct a tree with $N+2$ external legs. In
practice, we start from the root, which has a tree with $N+2$ external
legs attached to it. Then we proceed with branching the root into two
trees with $k+1$ and $N+2-k$ external legs (remember that we also
count the root leg), where $k$ is computed from Eq.~\rf{*k1}. At each
step we assign the number of external legs of the tree attached to
each link according to the same formula and we keep an ordered list of
the external legs of the whole tree. We add one such link to the list
whenever $k=1$.

Then we proceed to connect the external legs with a rainbow
diagram. This is possible since the total probability is the product
of the probability of constructing the rooted tree Eq.~\rf{*k1} and
the probability of constructing the corresponding rainbow diagram.
According to Eq.~\rf{xx2}, the probability of splitting a rainbow
diagram with $n$ lines into two rainbows with $k$ and $n-1-k$ lines is
given by:
\beq{*k2}
u(n,k) = \frac{R_k R_{n-k-1}}{R_n} = w(n+1,k+1)\, ,
\eeq
since $R_{n-1}=T_n$. In practice we start from the root leg of the
tree diagram and we split the rainbow containing $N/2+1$ lines into
two parts containing $k$ and $N/2-k$ lines. Then we can connect the
root leg with the appropriate member of the list containing the
external legs of the tree mentioned above and proceed accordingly
until we connect all external legs.

\subsection{The simulations}
\label{ss:ts}

The simulations are performed by generating a number of statistically
independent configurations using the algorithm mentioned above. We use
the high quality random number generator RANLUX \cite{ml,fj} whose
excellent statistical properties are due to its close relation to the
Kolmogorov K-system originally proposed by Savvidy
et.al. \cite{ss,ass} in 1986. We centered our effort for good
statistics on system sizes ranging from $2000${--}$256000$
triangles. The number of configurations obtained depends on the
lattice size and on the observable that we measure. We choose $20$
random vertices/triangles on each configuration in order to perform
correlation function measurements.  We need to collect more statistics
to test Eq.\ \rf{*77}, where we have between $4.2\times 10^6$ and
$1.6\times 10^6$ configurations.  For the $128K$ and $256K$ lattices
we have $6\times 10^5$ and $2\times 10^5$ configurations respectively.
It was possible to extract useful information about the short distance
behaviour of the two point function $n_N(r)$ by generating a smaller
number of configurations for system sizes having $512000${--}$8192000$
triangles. We got $10000$ configurations for $N=512000$ and $1024000$,
$4000$ for $N=2048000$ and $386$ for $N=8192000$ and we measured
$n_N(r)$ by choosing many more initial random
triangles/vertices. It is not possible to extract $d_H$ by
using finite size scaling with so low statistics.

In order to measure the moments $\la l^n \ra_{r,N}$ and their scaling
properties we need a factor of $10^2$ less configurations: We have
approximately $50000$ configurations for each lattice size.
Unfortunately, the computer effort for making the measurements is
comparable to the one needed to test Eq.\ \rf{*77} with enough
accuracy.

\begin{figure}[htb]
\centerline{\epsfxsize=4.0in \epsfysize=2.67in \epsfbox{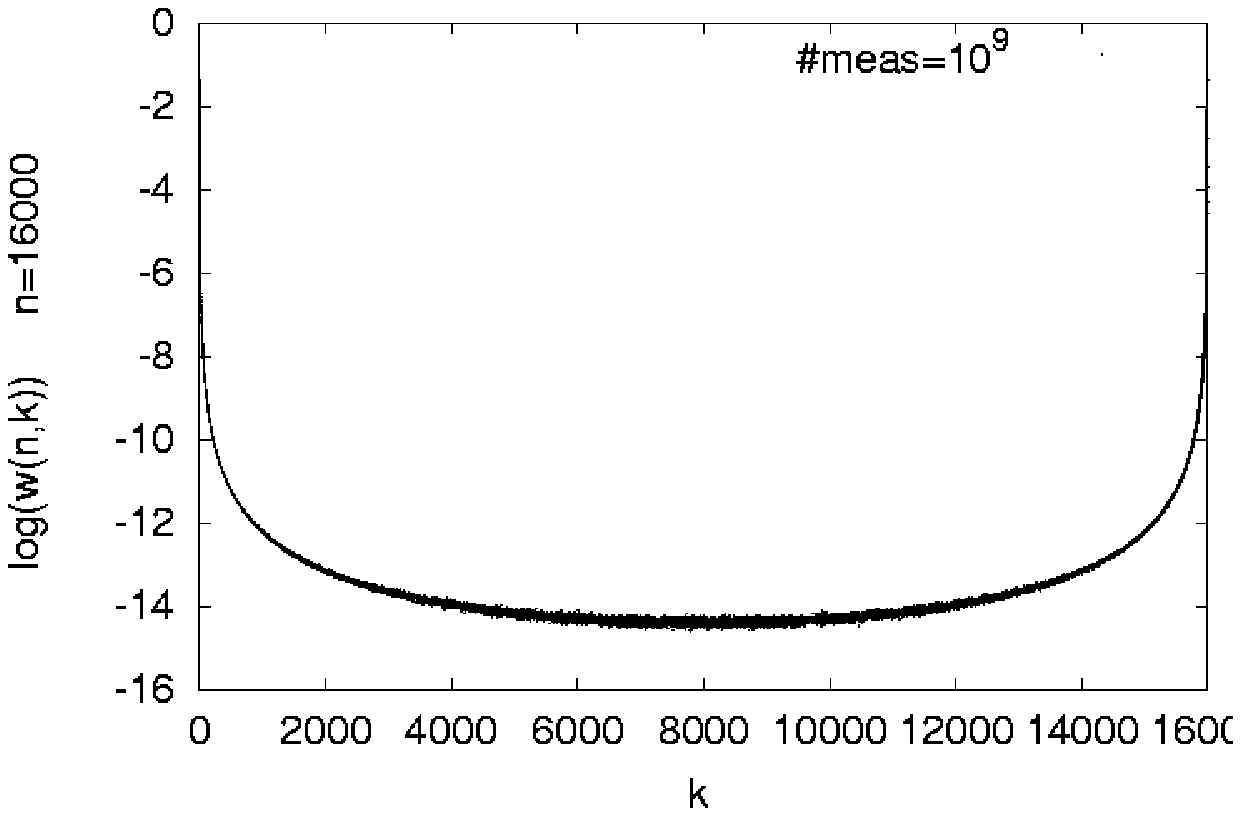}}
\centerline{\epsfxsize=4.0in \epsfysize=2.67in \epsfbox{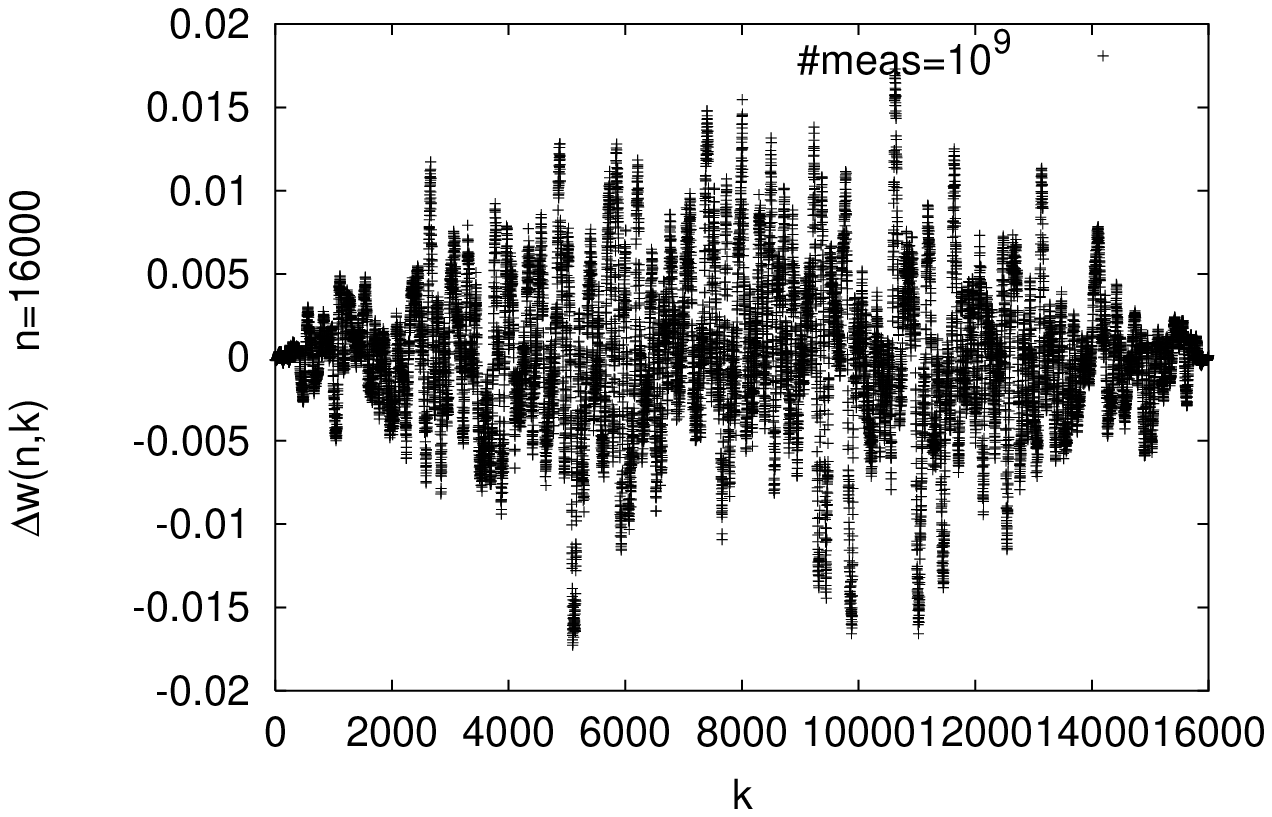}}
\caption{({\it a}) $w(n,k)$ for $n=16000$ plotted from Eq.~\rf{*k1}
and by application of our program routine ${\rm 10}^{\rm 9}$ times. 
({\it b}) The deviation $\Delta
w(n,k)=(w(n,k)_{\rm theoretical}-
w(n,k)_{\rm measured})/w(n,k)_{\rm theoretical}$
using the same parameters.}
\label{f:l}
\end{figure}

One subtle point in the simulations is the computation of the
branching numbers given by Eqs.~\rf{*k1} and \rf{*k2}. Given the
number $n$ we need to compute $k$. We do this by choosing a random
number $r$ in the interval $[0,1)$. Then, e.g. for the case of trees, 
we compute
\beq{*k3}
W(n,k) = \sum_{i=1}^k w(n,i)\, ,
\eeq
and we choose $k$ to be the integer such that $W(n,k)\ge r$. Using the
symmetry of $w(n,k)$ around $n/2$ we can substantially reduce the
computation time. Moreover $w(n,k)$ is best computed from the
recursive formula 
\beq{*k4}
w(n,k) = \frac{(2k-3)(n-k+1)}{k(2(n-k)-1)} \, w(n,k-1) \, ,
~~~~~
w(n,1) = \frac{n}{2(2n-3)} \, ,
\eeq
and extra care must be put in the computer program so that we obtain
the correct probabilities due to overflow. For this reason we have
tested the distributions of $k$ for given $n$ obtained from the
computer versus the theoretical value Eq.~\rf{*k1}. The results for
$n=16000$ are shown in fig.~\ref{f:l}(a) where we plot $w(n,k)$ from
Eq.~\rf{*k1} together with the results obtained from using our program
${\rm 10}^{\rm 9}$ times. A measure of the agreement is shown in
fig.~\ref{f:l}(b) where we plot the relative deviation $\Delta
w(n,k)=(w(n,k)_{\rm theoretical}-
w(n,k)_{\rm measured})/w(n,k)_{\rm theoretical}$. In the latter case, the
data is smoothed using Savitzky-Golay filters to interpolate the
nearest 100 points to a 4th order polynomial. As is seen, the
distribution is symmetrical and the relative errors average out to 0 very
nicely indicating that the deviation is pure statistical noise.
We have checked that $|\Delta w(n,k)| \sim N_{\rm measurements}^{-0.5}$.

\section{Numerical results}
\label{s:nr}

\subsection{The fractal dimension }
\label{ss:fd}

We have measured the fractal dimensions $d_H$ and $d_h$ in a number of
ways to be described in the following. Since we {\em are} considering
discretized, finite systems as approximations to continuum systems
(although we try of course to stay as close to continuum physics as
possible) it is important to use independent ways to approximate the
same continuum physical observable. This will often give a more
reliable idea of how close we are to the genuine continuum quantity
since it supplements the systematic large $N$ study of a single choice
of discretization of the continuum observable, where systematic
cancellation of fluctuations might sometimes underestimate the real
discrepancy between the measured quantity and the unknown value of the
continuum observable.

\subsubsection{Short distance behaviour}
\label{sss:sdb}

\begin{figure}[htb]
\centerline{\epsfxsize=4.0in\epsfysize=2.67in\epsfbox{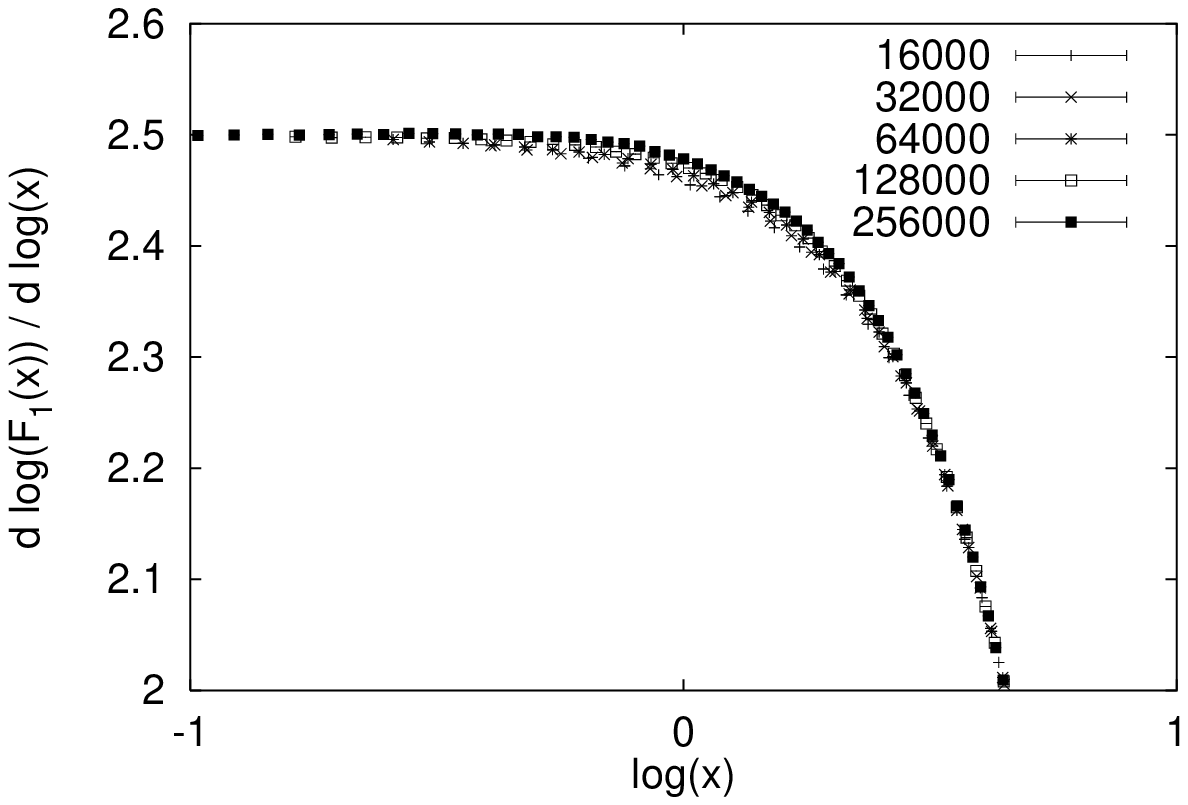}}
\centerline{\epsfxsize=4.0in\epsfysize=2.67in\epsfbox{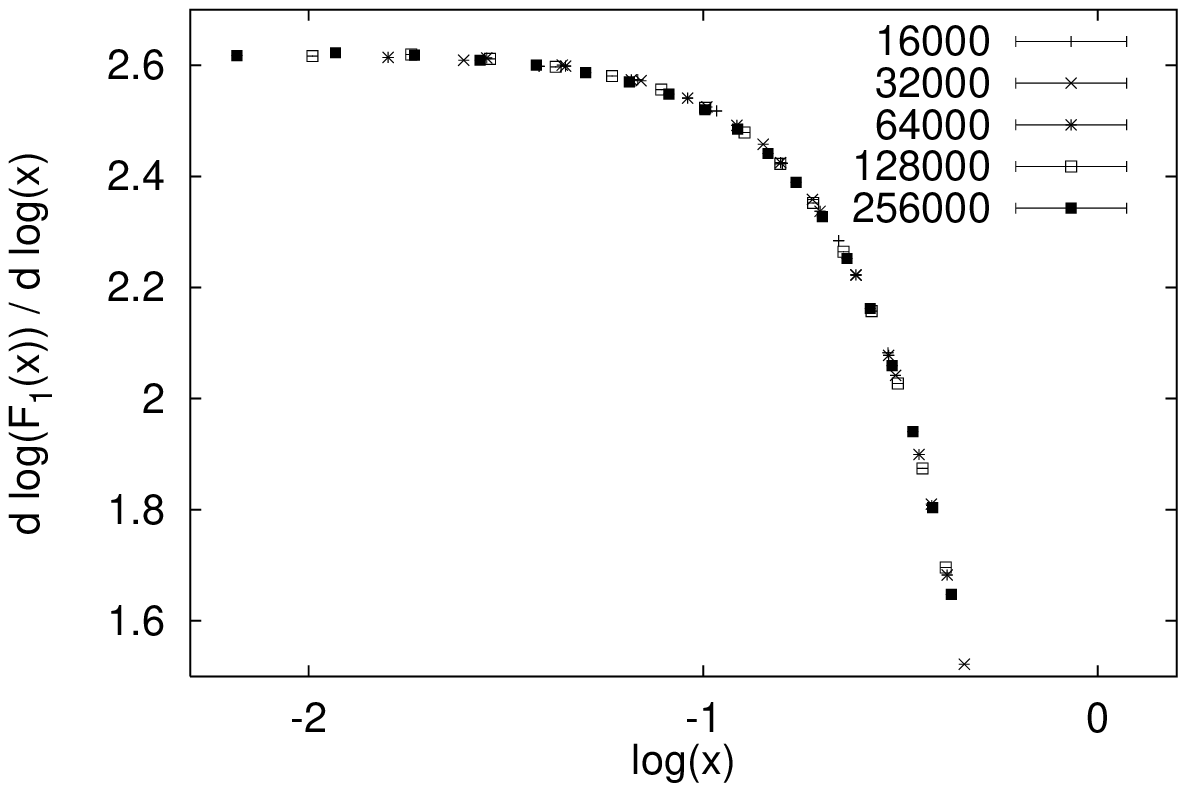}}
\caption{({\it a}) The small $x$ behaviour of the logarithmic
derivative of $n_N(r)$ where $r$ is the triangle distance. We use
$d_H=3.50$ and $a=4.11$.  ({\it b}) The same as in ({\it a}) but now
$n_N(r)$ is defined in terms of link distance, $d_H=3.62$ and
$a=0.52$.}
\label{f:5}
\end{figure}

One can try to use directly the short distance behaviour \rf{*76} of
$n_N(r)$ to extract $d_h$. This was the method used in the pioneering
work \cite{kketal}. The problematic aspect of the method is to what
extend one has to take seriously the natural requirement $1 \ll r \ll
N^{1/d_H}$. We know now from the exact solution of pure
two-dimensional quantum gravity that both limits have to be respected
quite seriously, but that a so-called shift $r \to r+a$ helps to
almost remove the requirement for the lower limit \cite{ajw}. Later we
will discuss various theoretical and ``phenomenological'' motivations
for this shift. Presently, let us just use it as an additional fit
parameter.  In table \ref{t:2} and table \ref{t:3} we have shown the
results of a fit of the form
\beq{*83}
n_N(r) = C \, (r+a)^{d_h-1},
\eeq
for various cuts of $r_{\rm min} \leq r \leq r_{\rm max}$.
We see here a  discrepancy between  $d_h$ obtained from the link-distance
measurements and the triangle-distance measurements, respectively. 
If there {\em is} a discrepancy between the measured $d_h$ for 
triangle and link distance for finite $N$ one would expect it to  
be ``maximal'' when we concentrate on small $r$, like here. For 
small $r$ the triangle distance is quite rigid in the sense that 
each triangle has at most three neighbours, 
while  a vertex can have any number of neighbouring
vertices. This rigidity is also reflected in the fact that 
we have to use a much larger shift $a$ for triangle-distances.
There is a weak tendency for the link--$d_h$ to decrease with $N$ while
the triangle--$d_h$ shows a similar weak tendency to increase with $N$.
From this one would conservatively estimate, assuming that 
they have a common large $N$ limit, that
\beq{*84}
3.53 < d_h < 3.60.
\eeq
In figs.\ \ref{f:5}a and \ref{f:5}b we have shown the behaviour 
of short distance behaviour of $n_N(r)$ for the 
two distance measures and the whole range of $N$. 

We will now show that the discrepancy between link--$d_h$ and 
triangle--$d_h$ indeed decreases when we use the finite size 
scaling \rf{*77} to invoke the whole range of $r$ and also that 
we can get a much better determination of $d_h$.

\subsubsection{Collapse of distributions}

Before using the finite size scaling relation \rf{*77}
let us motivate the use of the ``shift'' $a$ which is 
essential for obtaining high precision results. 
The need of this parameter is well known from
earlier studies of conformal field theories with $c \geq 0$ coupled 
to quantum gravity \cite{ajw,ak}. One obvious,
``phenomenological'', motivation for this shift parameter (which 
we have just copied from standard finite size scaling theory) is 
as follows: for finite $N$ we expect some discrepancy compared 
to continuum results, typically parameterised by ``the number of 
points'' $L$ corresponding to the linear size of the system.
In particular we can write
\beq{*80}
x= \frac{R}{V^{1/d_H}} \sim   \frac{r}{N^{1/d_H}} + 
\frac{{\rm const.}}{L}+ \frac{{\rm const.}}{L^2} + \cdots,
\eeq  
or, since $N^{1/d_H}$ is precisely a typical measure for 
the linear extension of the system:
\beq{*81}
x  =  \frac{R}{V^{1/d_H}} 
\sim  \frac{r+a}{N^{1/d_H}} 
      + \frac{{\rm const.}}{(N^{1/d_H})^2} + \cdots.
\eeq
The parameter $a$, which is considered a {\em shift in r}, incorporates
the first order correction. Below we  derive a theoretical 
value for $a$ in the case of pure gravity and where we use 
triangle distance. However, here we consider it as a purely 
phenomenological parameter, which in principle can be different for 
different scaling variables and {\em will} be different if we 
use different distance measures (viz.\ link and triangle distances).

The raw measurements of $n_N(r)$ produces distributions with
$N$ ranging from $1K$ to $512K$. We can now try to fit them 
to the scaling ansatz \rf{*77}. We have two parameters 
available, $d_H$ and $a$.

The collapse of the distributions is performed by using two different
methods. The first one is identical to the one proposed in
\cite{suracuse} and used with great success also in \cite{ak,us}. One
makes a non linear fit of the form $p_n(x){\rm e}^{-m x}$, where
$p_n(x)$ is a polynomial of order $n$ in $x$, to the rescaled
according to Eq.~\rf{*77} distributions $n_N(r)$ for a given set of
lattice sizes $\{ N \}$. The $\chi^2(a,d_H)$ per degree of freedom is
computed for a given set of parameters $a$ and $d_H$. For each value
of the shift $a$, the optimal value of $d_H$ is computed from the
position of $\chi^2_{\rm min}(a)$.

We also used a second method which has the advantage of being much
faster. We should also note that it is also quite successful even for
very small lattices ($N<500$) where the fits used in the first method
fail to yield reasonable results. For a given set of lattice sizes $\{
N \}$ and parameters $(a,d_H)$ we compute a cubic spline interpolation
to the rescaled distribution for each value of $N$. $\chi^2(a,d_H)$ is
computed by adding in quadrature the distances of each point of the
other distributions from the interpolation function reweighted by their
errors and properly normalised to correspond to a $\chi^2$ per degree
of freedom. The best values for $d_H$ computed this way are identical
to the ones computed using the first method, although $\chi^2(a,d_H)$ is
slightly smaller and steeper yielding smaller errorbars ($\approx$
10--30\%) for the computed quantities. In this paper we report the
larger errors computed from the first method.

\begin{figure}[htb]
\centerline{\epsfxsize=4.0in \epsfysize=2.67in \epsfbox{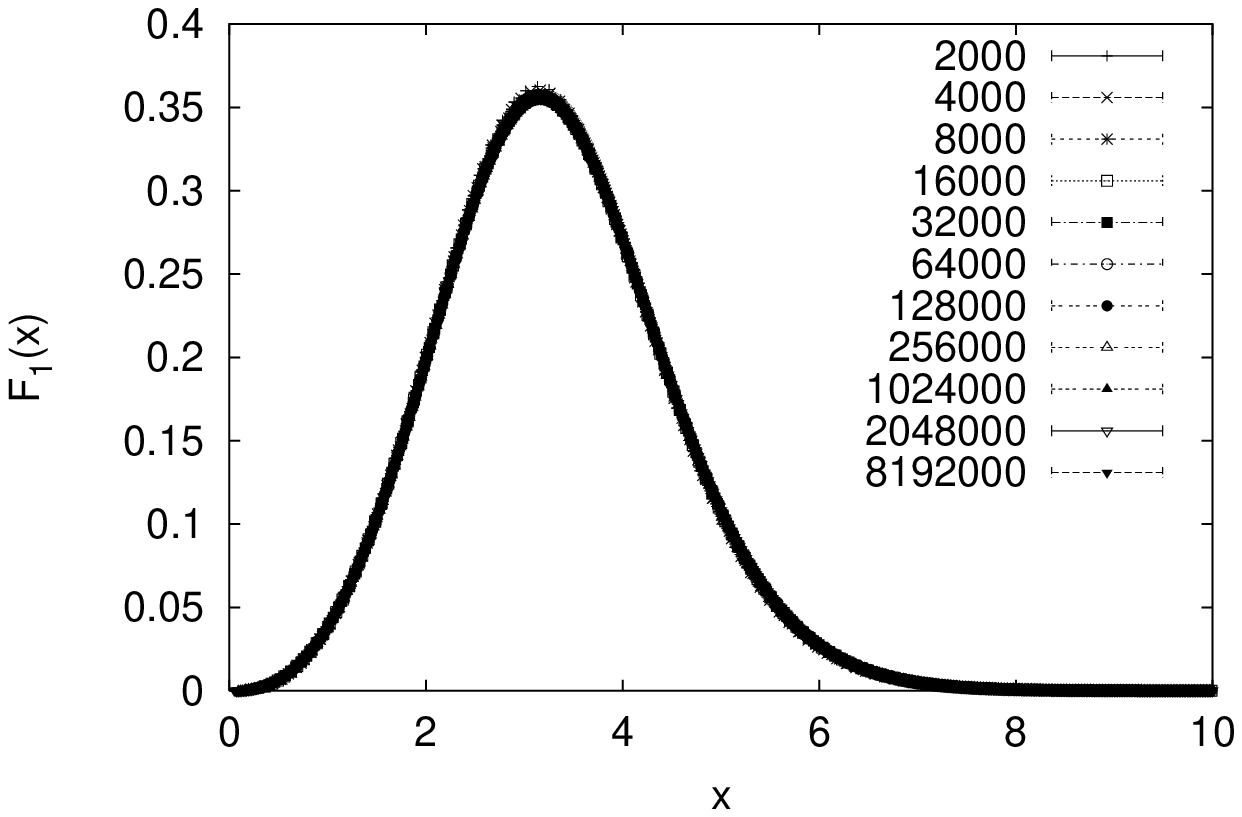}}
\centerline{\epsfxsize=4.0in \epsfysize=2.67in \epsfbox{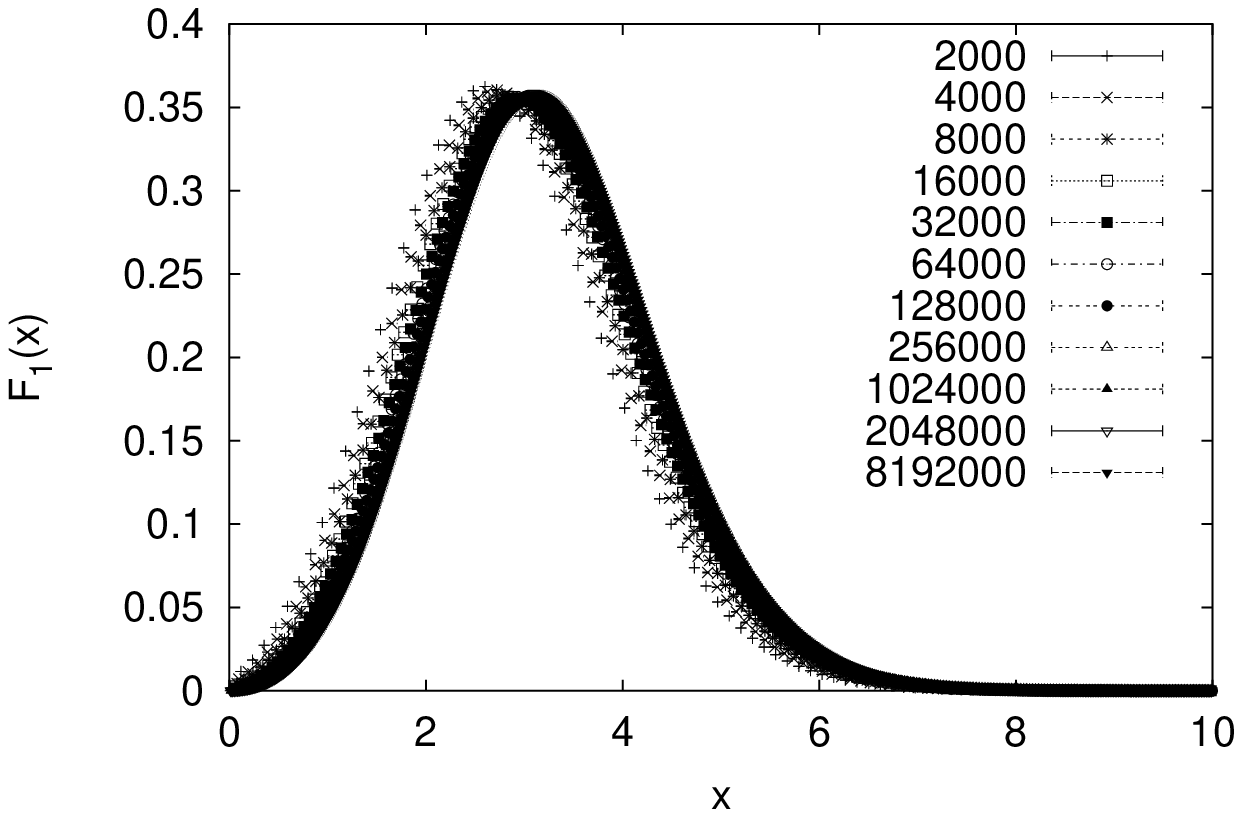}}
\caption{({\it a}) The $n_N(r)$ distributions defined in terms of
triangle distance rescaled according to Eq.~\protect\rf{*77} 
using $d_H=3.56$ and
$a=4.50$. ({\it b}) The same as in ({\it a}) but now by setting
$a=0$.}
\label{f:1}
\end{figure}

\begin{figure}[htb]
\centerline{\epsfxsize=4.0in \epsfysize=2.67in \epsfbox{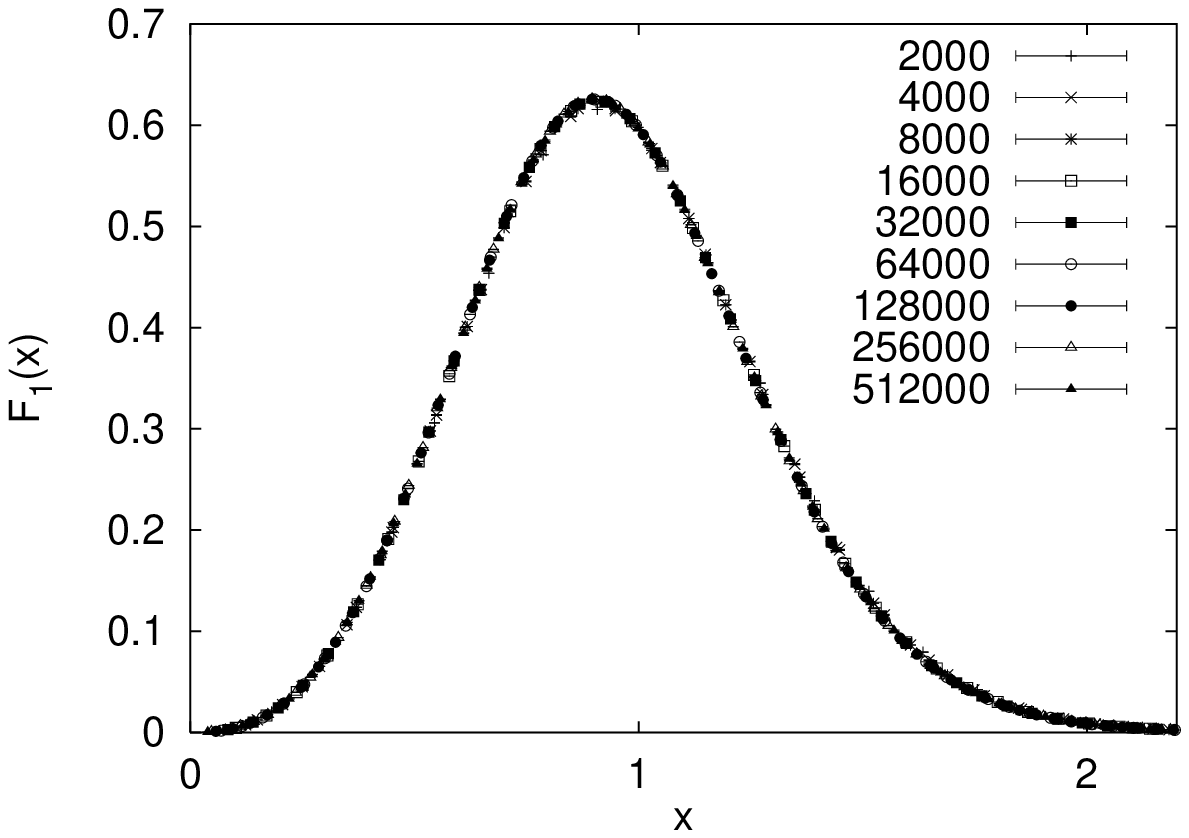}}
\centerline{\epsfxsize=4.0in \epsfysize=2.67in \epsfbox{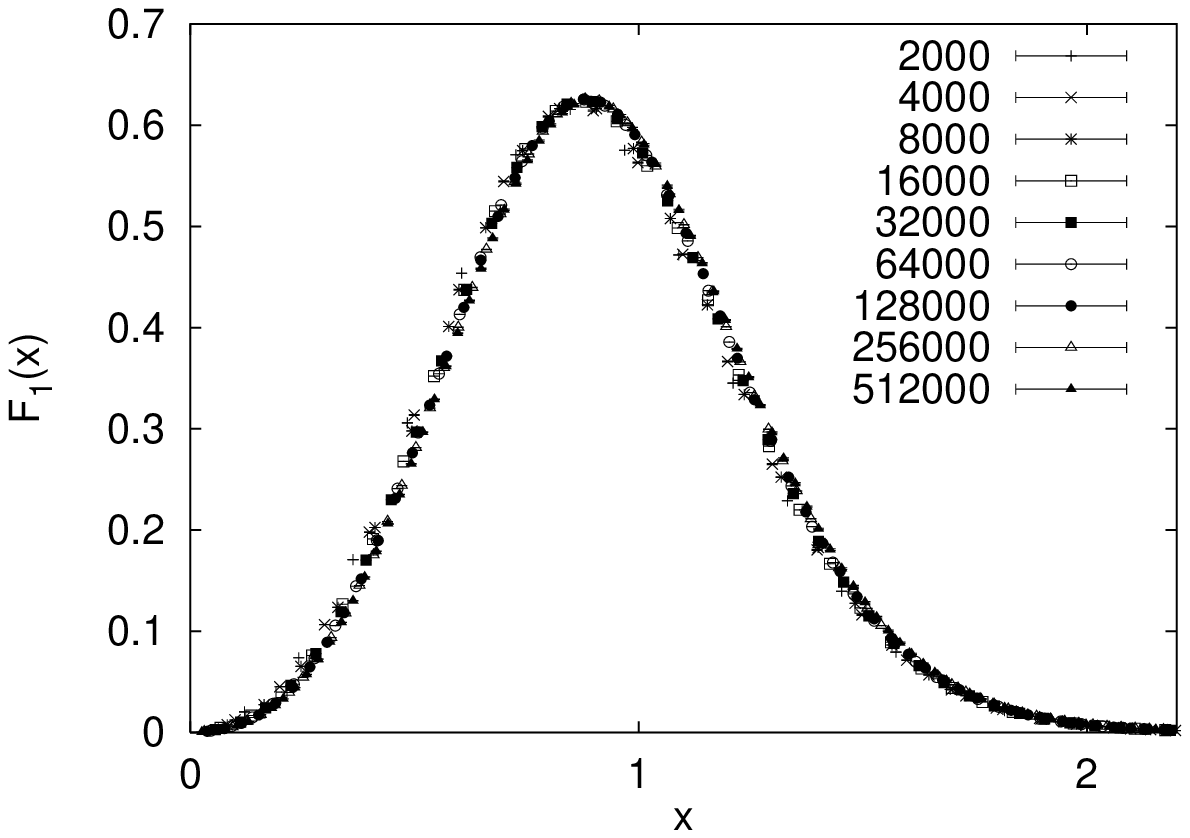}}
\caption{({\it a}) The $n_N(r)$ distributions defined in terms of
link distance rescaled according to Eq.~\protect\rf{*77} using $d_H=3.60$ and
$a=0.50$. ({\it b}) The same as in ({\it a}) but now by setting
$a=0$.}
\label{f:2}
\end{figure}

\begin{figure}[htb]
\centerline{\epsfxsize=4.0in \epsfysize=2.67in \epsfbox{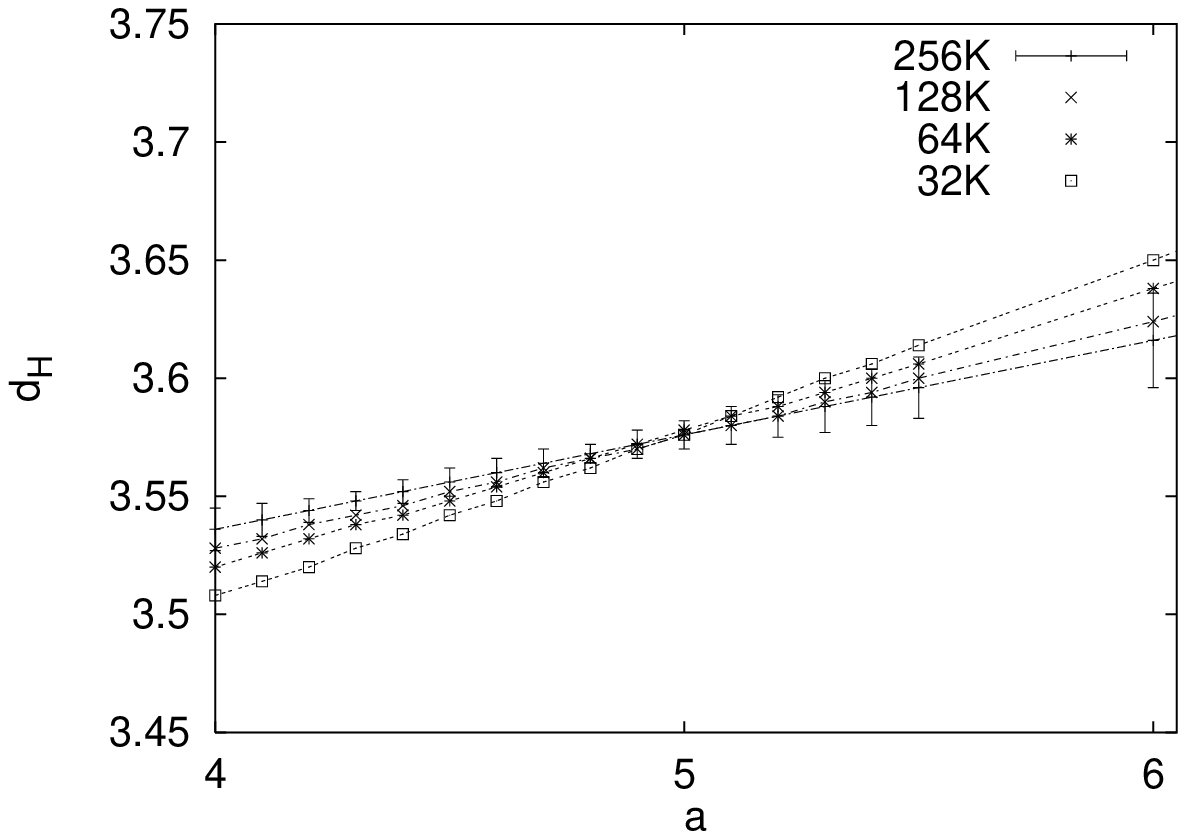}}
\centerline{\epsfxsize=4.0in \epsfysize=2.67in \epsfbox{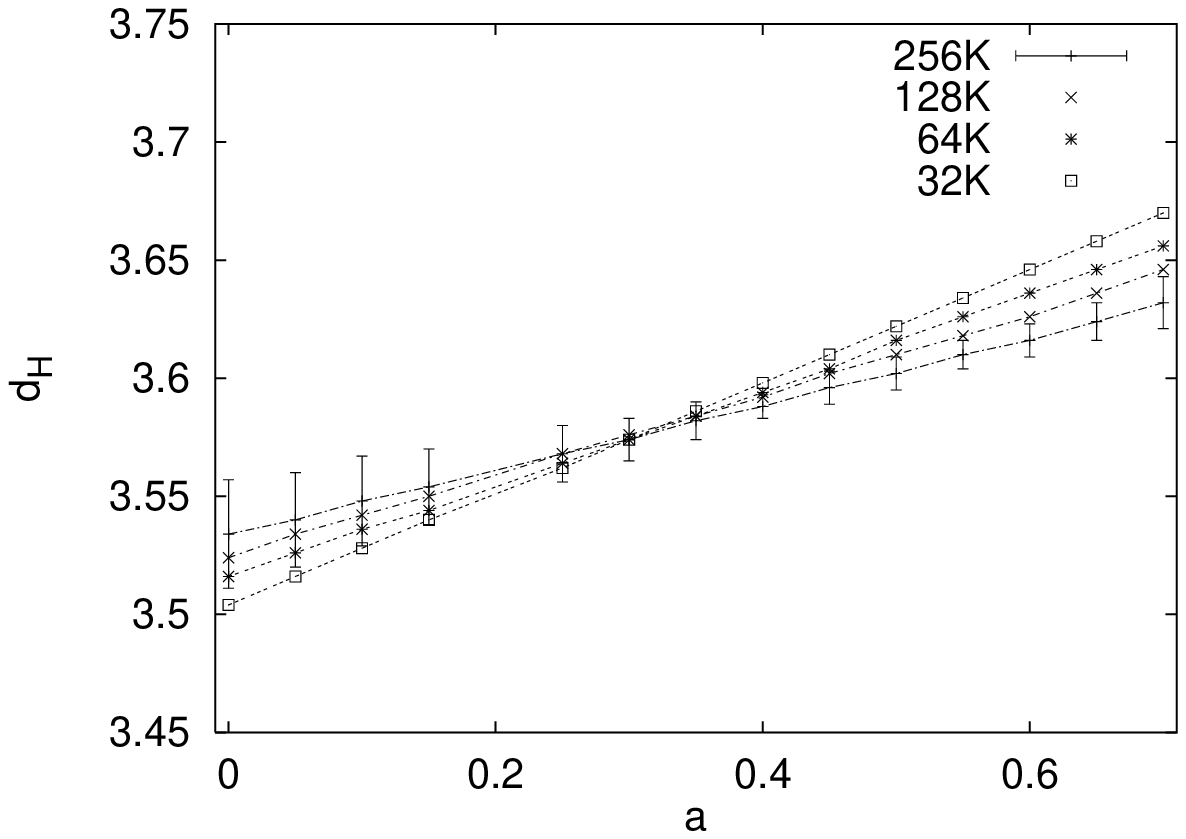}}
\caption{({\it a}) $d_H(a)$ computed from collapsing the $n_N(r)$
distributions defined in terms of triangle distance. We collapse the
distributions pairwise and in the plot we indicate the largest
lattice. For clarity, errors computed from $\chi^2$ are displayed
only for the largest lattice. ({\it b}) The same as in ({\it a}) but
now  $n_N(r)$ is defined in terms of link distance.}
\label{f:3}
\end{figure}

In table \ref{t:1} we have shown the results of the best fits
both for the link-distance distribution and the triangle-distance 
distribution. The fits are divided into groups: the upper one 
determine $d_H$ and $a$ by comparing distributions $n_N(r)$ for 
successive pairs of $N$'s ($16K$ and $32K$, $32K$ and $64K$, etc.) 
In this way the $(d_H,a)$ dependence on $N$ becomes clear. 
The middle and lower groups joins three, 
respectively four successive $N$'s in the determination
of $d_H$ and $a$. 
In particular for the top group we see that $d_H$ has a
clear dependence on $N$: If we use link distances $d_H(N)$ is
systematically decreasing, while $d_H(N)$ is systematically increasing 
if we use triangle distances. Under the assumption that the trend will
continue for larger systems and that the two distance measures
are proportional to the genuine continuum distance in the scaling 
limit we conclude that 
\beq{*82}
3.55 < d_H < 3.61.
\eeq
These values of $d_H$ (and the appropriate values of $a$) actually
yield excellent finite size scaling for the whole range of $N$ is
illustrated in fig.\ \ref{f:1}a where we have shown $F^{(N)}_1(x)$ for
the whole range of $N$ from $2K-256K$. The importance of the inclusion
of the shift $a$ if we use the triangle distance is illustrated in
fig.\ \ref{f:1}b where we have plotted the functions $F^{(N)}_1(x)$
obtained from $n_N(r)$ with the same choice of $d_H$ but with $a =0$.
The same plots are shown in fig.\ \ref{f:2}a-\ref{f:2}b in the case
where we use the link distance. Here $a$ is smaller and not of the
same visual importance as for triangle distances. However, in the
actual fits they play an important role for the link distances as well
if we want to extract consistent values of $d_H$ with acceptable
$\chi^2$ values for the fits. In fig.~\ref{f:3} we see that the value
of $d_H$ depends strongly on $a$.
  
\begin{figure}[htb]
\centerline{\epsfxsize=4.0in \epsfysize=2.67in \epsfbox{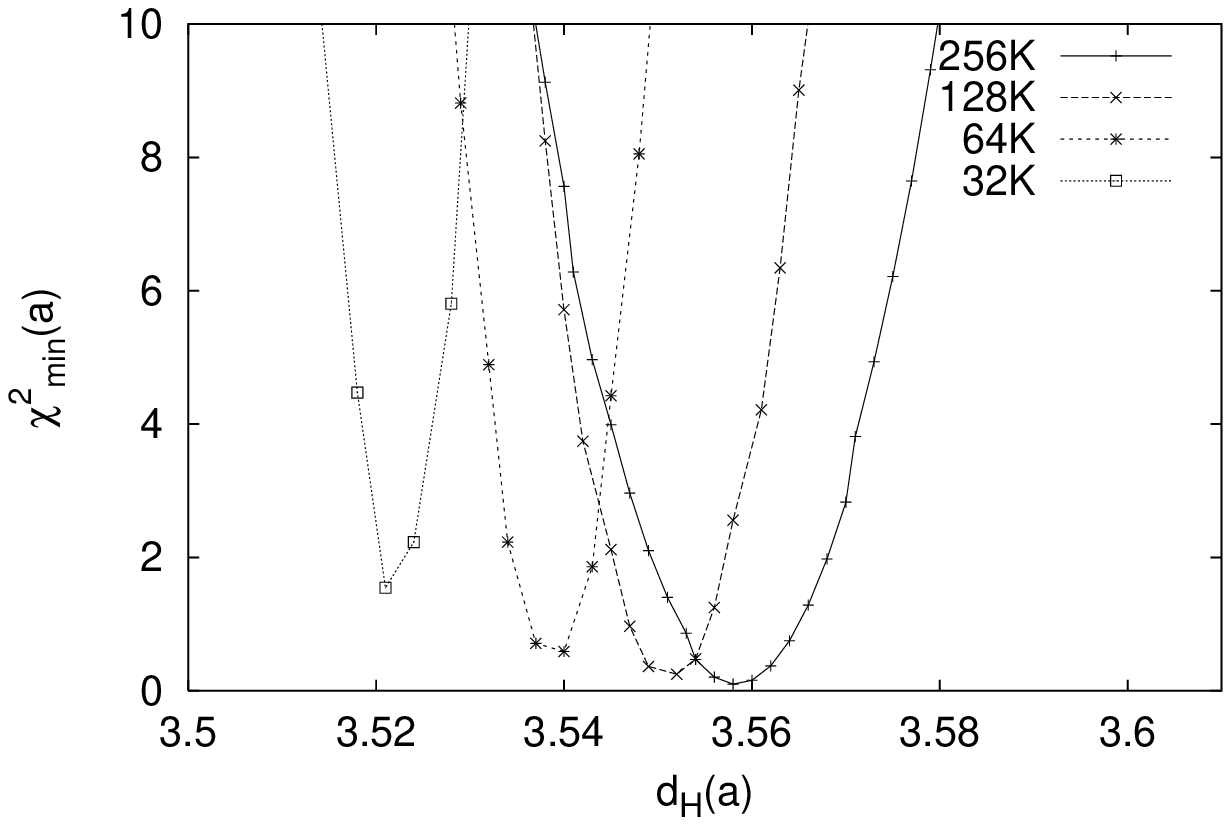}}
\centerline{\epsfxsize=4.0in \epsfysize=2.67in \epsfbox{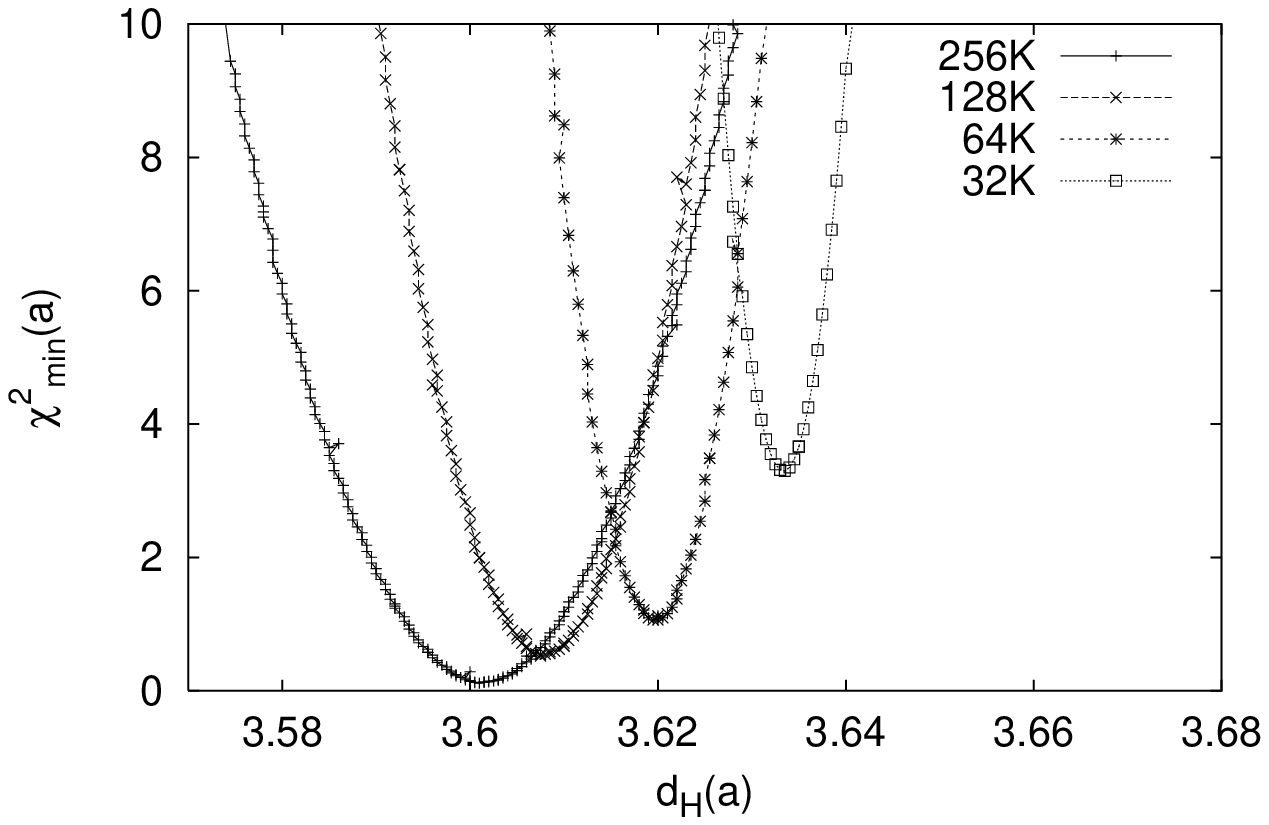}}
\caption{({\it a}) $d_H$ computed from $\chi^2_{\rm min}$ by collapsing
the $n_N(r)$ distributions defined in terms of triangle
distance. Each point on the graph is $d_H(a)$ (see
Fig.~\protect\ref{f:3}) and is plotted versus the corresponding
$\chi^2_{\rm min}(a)$. ({\it b}) The same as in ({\it a}) but now
$n_N(r)$ is defined in terms of link distance.}
\label{f:4}
\end{figure}

However, we can use the systematic behaviour of $d_H$ as a function of
$N$ {\em and} the shift parameter $a$ to obtain a better estimate of
$d_H$ than the one provided by \rf{*82}.  In fig.\ \ref{f:4} we have
shown the complete $\chi^2_{\rm min}(a)$ fit which was used in the top
group of table \ref{t:1}. Each point on the two subfigures represents
a specific choice of the shift $a$.  For a given $N$ we now find the
value $d_H(a,N)$ which minimises the $\chi^2$ of the collapse of the
$n_N(r)$ and $n_{N/2}(r)$.  In table \ref{t:1} we recorded just the
minimum as a function of $a$ and the change of the minimum as a
function of $N$ is seen quite clearly in fig.\ \ref{f:4}. However, let
us for each $N$ plot the $d_H(a,N)$ of fig.\ \ref{f:4} as a function
of the shift $a$.  This is shown in fig.\ \ref{f:3}. We get a number
of straight lines which with very good accuracy intersect for one
value of $a_0$.  The simplest phenomenological explanation of this
fact is that $a_0$ is the correct value from \rf{*82}, since this
implies that
\beq{*85}
d_H(a,N) = d_H(a_0,\infty)+ 
\left[\frac{\prt d_H(a_0,\infty)}{\prt a} + 
\frac{\prt^2 d_H(a_0,\infty)}{\prt a_0 \prt (1/N)} \, \frac{1}{N}\right]
(a-a_0)+ \cdots.
\eeq
From this argument the correct infinite volume limit of $d_H$ is 
determined from the figures up to the accuracy with which the 
curves actually cross in a single point. This interpretation 
confirmed by the fact that both the triangle and link $d_H(a,N)$ 
curves cross at the same value of $d_H$, but for very different $a_0$.

\vspace{6pt}
\noindent
We conclude from the data (see figure) that 
\beq{*86}
d_H = 3.57\pm 0.01 \, .
\eeq

\subsubsection{Average radius}

\begin{figure}[htb]
\centerline{\epsfxsize=4.0in \epsfysize=2.67in \epsfbox{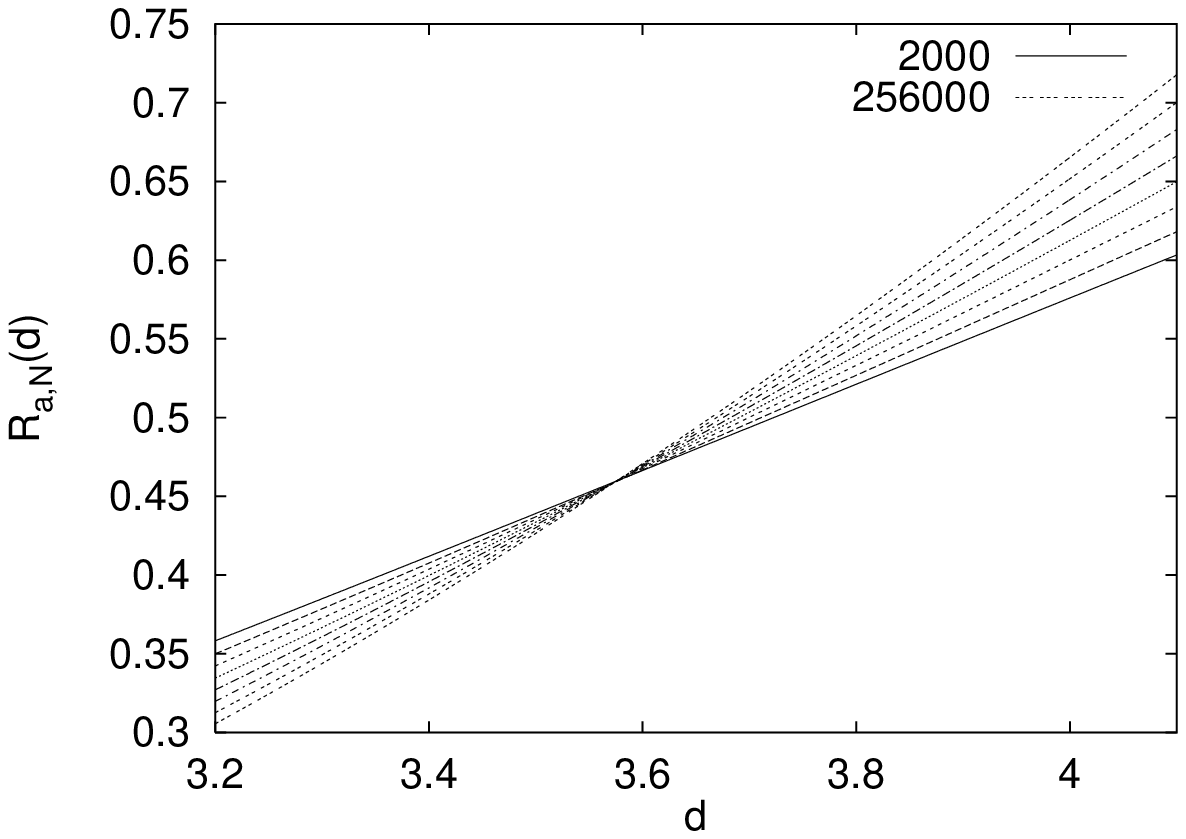}}
\centerline{\epsfxsize=4.0in \epsfysize=2.67in \epsfbox{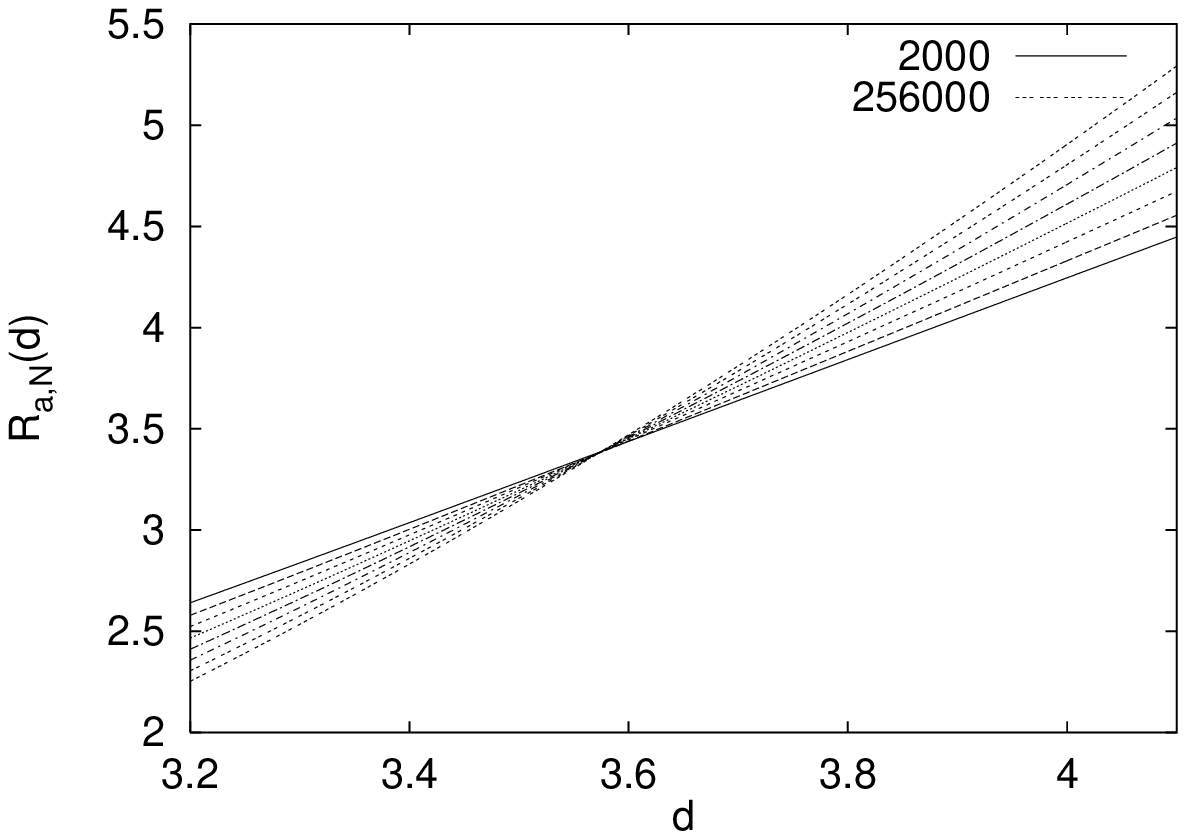}}
\caption{ ({\it a}) The functions $R_{a,N}(d)$, defined in terms of
link distance, for $N=2K,4K,\ldots,256K$ and $a=0.130$.  ({\it b})
Same as in ({\it a}) but when $R_{a,N}(d)$ is defined in terms of
triangle distance and $a=5.0$.}
\label{f:11}
\end{figure}

\begin{figure}[htb]
\centerline{\epsfxsize=4.0in \epsfysize=2.67in \epsfbox{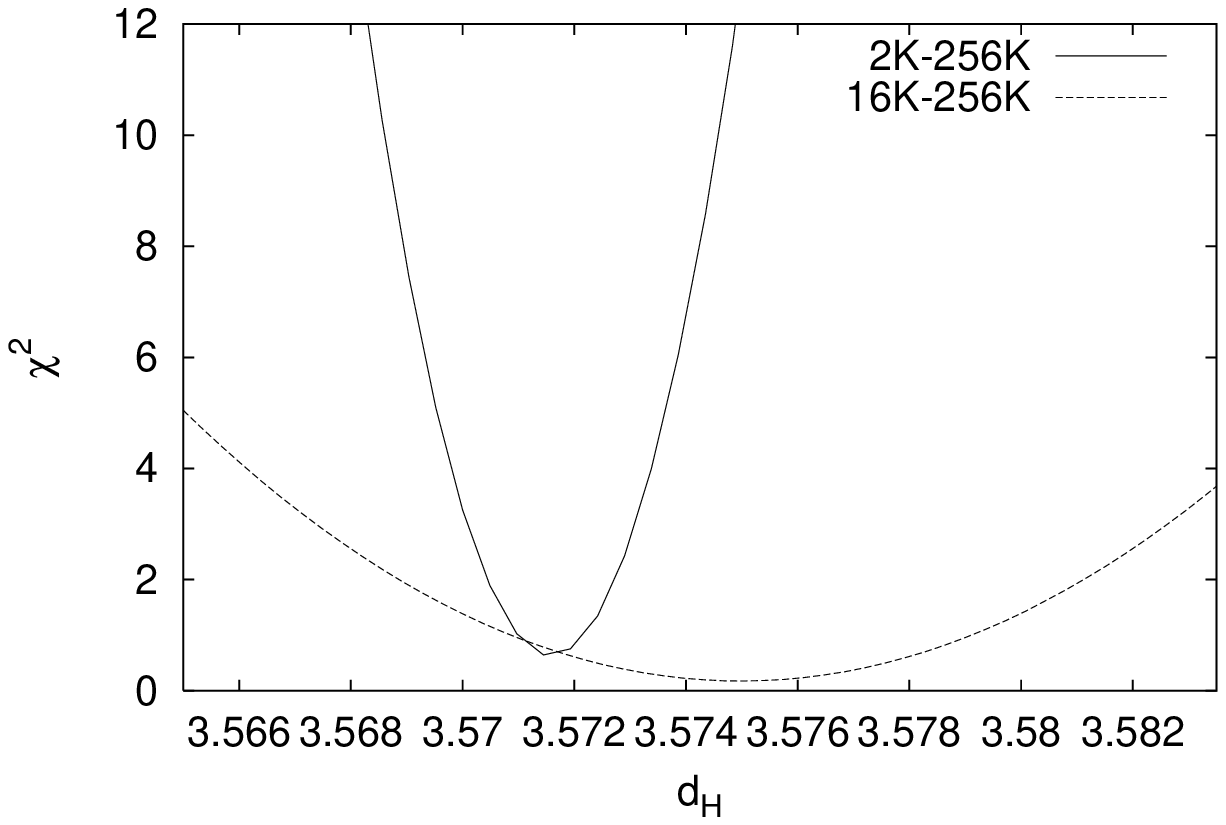}}
\centerline{\epsfxsize=4.0in \epsfysize=2.67in \epsfbox{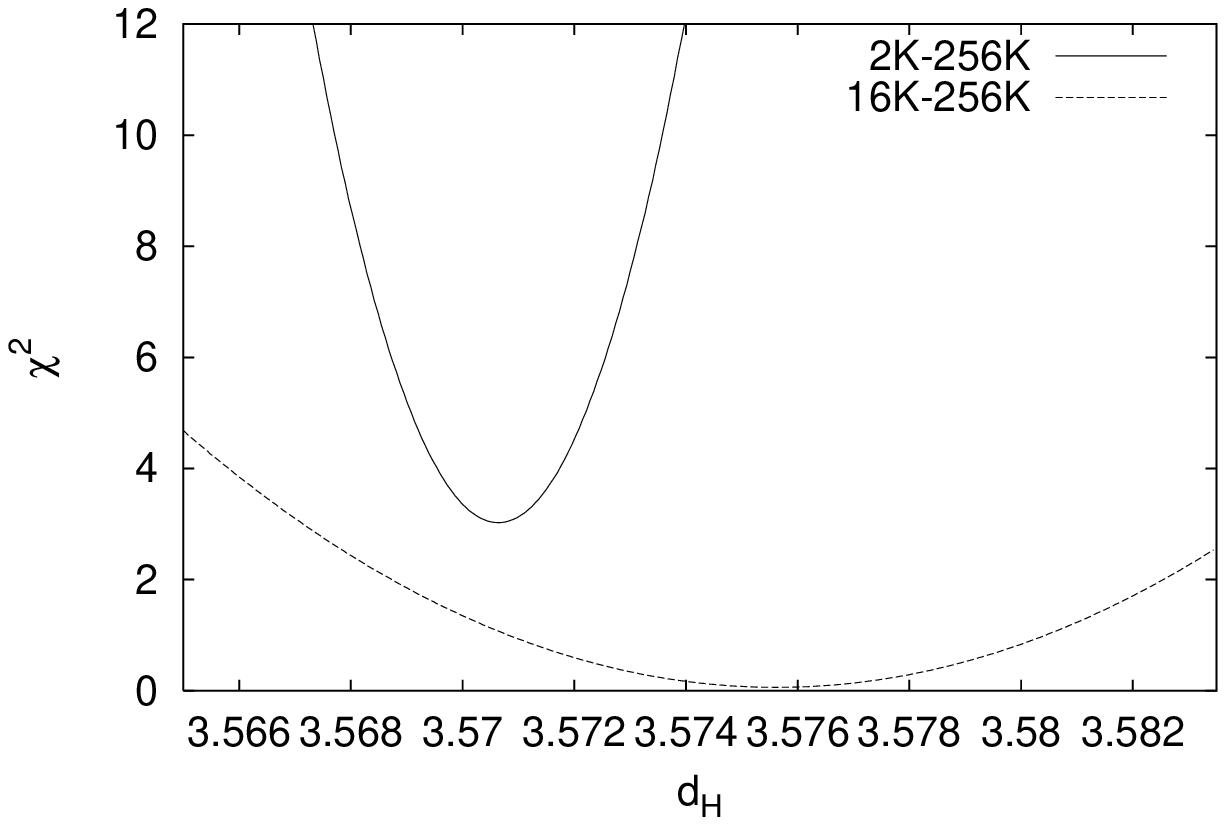}}
\caption{ ({\it a}) $\chi^2(a)$, defined by Eq.~\protect\rf{*18a} 
for two sets of
$N_i$'s when $R_{a,N}(d)$ is defined in terms of link distance.  ({\it
b}) Same as in ({\it a}) but when $R_{a,N}(d)$ is defined in terms of
triangle distance.}
\label{f:12}
\end{figure}

In this section we will use the average radius of a universe to extract
the intrinsic Hausdorff or fractal dimension.
From the definition of $n_N(r)$ the average radius of universes 
with volume $N$ is
\beq{*13a}
\la r \ra_N \equiv \frac{1}{N}\sum_{r=0}^\infty r \; n_N(r)  
\sim  N^{1/d_H}\, .
\eeq
Obviously, \rf{*13a} could itself serve as a natural definition of $d_H$. 
By measuring $n_N(r)$ we can record $\la r \ra_N$
as a function of $N$ and hence determine $d_H$.
As usual we have to introduce the shift $a$ in order to 
account for lowest order discretization effects.
 Now, let us define  
\beq{*15a}
R_{a,N}(d) = \frac{\la r + a \ra_{N}}{N^{1/d_H}}\, .
\eeq
We determine the values 
of $a$ and $d_H$ in the following way: first we measure $\la r \ra_{N_i}$
for a certain number of different volumes $N_i$ of the universes, 
ranging from $N=2K$ to $N=256K$. 
For a given $a$ we choose, for {\it each couple} $N_i,N_j$ of $N$'s, 
the $d_H^{ij}$ such that 
\beq{*16a}
R_{a,N_i}(d_H^{ij}) = R_{a,N_j}(d_H^{ij}) \, .
\eeq   
For this choice of $N_i,N_j$ we bin the data and estimate an error 
$\del d_H^{ij}$. Then we determine the average 
\beq{*17a}
\bar{d}_H  =  \frac{1}{{\rm \#~pairs}} \sum_{i\neq j} d_H^{ij} \, ,
\eeq
and compute 
\beq{*18a}
\chi^2(a) = \frac{1}{{\rm \#~pairs}} 
\sum_{i\neq j} \frac{(d_H^{ij} - \bar{d}_H)^2}{(\del d_H^{ij})^2}\, .
\eeq
The preferred pair $(a,d_H(a))$ is determined by the minimum of
$\chi^2(a)$.  This method works quite impressively.  In fig.\
\ref{f:11}a and fig.\ \ref{f:11}b we have shown the intersection of
the curves $R_{a,N}(d)$ as a function of $d$ for the optimal choice of
$a$ for link--distance and triangle--distance measurements,
respectively.  The important points are that in both cases there
exists a value of $a$ where all the curves intersect with high
precision, and that the range of $a$ where $\chi^2(a)$ is acceptably
small, i.e.\ $O(1)$, is quite small. Hence $d_H$ is determined with
high precision. In Fig.\ \ref{f:12}a and \ref{f:12}b we show
$\chi^2(a)$ for link--distance and triangle distance measurements,
respectively.  In this way we get
\beq{*19a}
d_H(a_m) = 3.573\pm 0.005\,
\eeq
both from the link--distance measurements and the triangle--distance 
measurements. The values of $a_m$ are 
\beq{*19b}
a_m= 0.13\pm 0.01\, ,~~~~~{\rm and}~~~~~a_m =5.00 \pm 0.05,
\eeq  
for the link--distance and triangle--distance, respectively.
In table \ref{t:6} we have listed the determination of 
$(d_H(a_m),a_m)$ for various cuts in the lower values of 
$N$. The constancy and consistency of the results are 
truly remarkable.

In \rf{*19a} we have estimated the error as follows. 
Define an interval of acceptance $[a_{\rm min},a_{\rm max}]$ of $a$ 
by demanding that $\chi^2(a) < 2 \chi^*$ 
where $\chi^* = {\rm max}\{1 , \chi^2(a_m)\}$
and find the variation of $d(a)$ in this interval.
After this we repeat the whole procedure by making various cuts in the 
pairs of $N_i$'s included in \rf{*17a} and \rf{*18a}, 
discarding successively the smallest $N_i$'s. 

\vspace{6pt}
\noindent
We note that the values \rf{*19a} are in perfect agreement with \rf{*86}.
We have now four independent measurements (two different methods, two
different definitions of length) which all give $d_H=3.57\pm 0.01$.

\subsection{Boundaries}

\begin{figure}[htb]
\centerline{\epsfxsize=4.0in\epsfysize=2.67in
\epsfbox{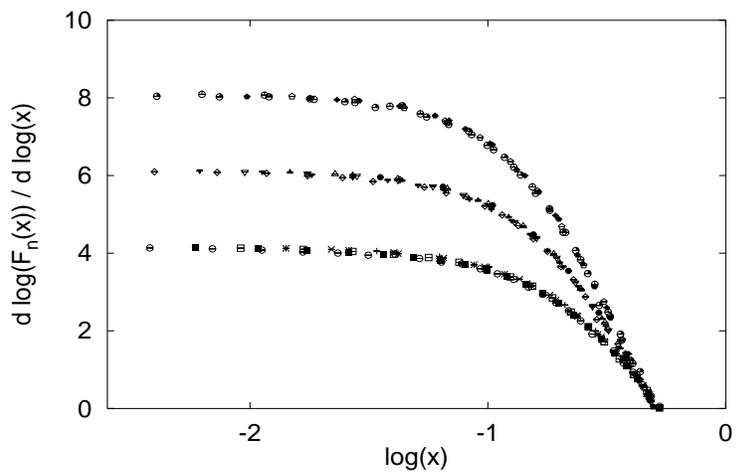}}
\caption{The small $x$ behaviour of the logarithmic derivative for
$\vev{l^n}_{r,N}$ for $n=2$, $3$, $4$, $N=16000${--}$512000$ using $a=0.32$
and $d_H=3.63,3.65$ and $3.66$ respectively.}
\label{f:8}
\end{figure}

We now turn to the measurements of $\la l^n \ra_{r,N}$. 
These observables are constructed from $\rho_N (l,r)$, which 
can readily be measured in the simulations. 
We recall the situation in  pure two-dimensional 
quantum gravity which can be solved explicitly and where 
$d_h=4$. In this case we have (for small $r$):
\beq{*90}
\la l \ra_{r,N}^{c=0} \sim r^3,~~~~\la l^n \ra_{r,N}^{c=0} \sim r^{2n},~~n>1.
\eeq
Since $d_h = 4$ in this case, i.e.\ $\la r^{2n} \ra \sim N^{n/2}$, we obtain
\beq{*91}
\la l^n\ra_{r,N}^{c=0} \sim \Bigl(\sqrt{N}\Bigr)^n,~~~n >1,
\eeq
as one naively would have expected for a smooth two-dimensional world.
Only the first moment behaves anomalous, as it {\em has} to do if $d_h \neq 2$.
From these $c=0$ considerations it is unclear what to expect for $c=-2$
for the higher moments. If ${\rm dim}[ \, N \, ]={\rm dim}[ \, l^2 \, ]$, 
then from scaling arguments, we expect
\beq{*20a}
\la l^n \ra_{r,N}  \sim  N^{n/2} \tilde{F}_n (x)\, ,
~~~~~x = \frac{r}{N^{1/d_H}}\, .
\eeq
However, our measurements are consistent with 
the following scaling relations 
\beq{*21a}
\la l^n \ra_{r,N}  \sim  N^{2n/d_H} F_n(x),
~~~~~{\rm for}~~n \geq 2\, , 
\eeq
which implies that ${\rm dim}[ \, l^n \, ]={\rm dim}[ \, r^{2n} \, ]$
for $n >1$. Eq.\ \rf{*21a} indicates that we have 
\beq{*22a}
\la l^n \ra_{r,N} \sim r^{2n}
~~~~~{\rm for}~~1 \ll r \ll N^{1/d_H},~n \geq 2\, .
\eeq
We have shown this relation in fig.\ \ref{f:8} for $n=2$, $3$ and $4$ 
for the case where we use link distances. It is remarkably well satisfied.
In table \ref{t:5} we have shown the more detailed result of this 
short distance analysis. It should be compared to the short distance 
analysis presented in table \ref{t:3} (and fig.\ \ref{f:5}b)
for the first moment. The behaviour of the exponents $2n$ for the
higher moments as functions of $N$ are clearly more consistent
and systematic compared with the behaviour of $d_H(N)$ for the first 
moment. In addition one can use smaller values of $r$ and there is not 
the same crucial dependence on the shift $a$ as for the first moment.
All this points consistently to smaller short-distance discretization
effects for the higher moments. Note also that the value of the shift $a$
is different.

\begin{figure}[htb]
\centerline{\epsfxsize=4.0in\epsfysize=2.67in\epsfbox{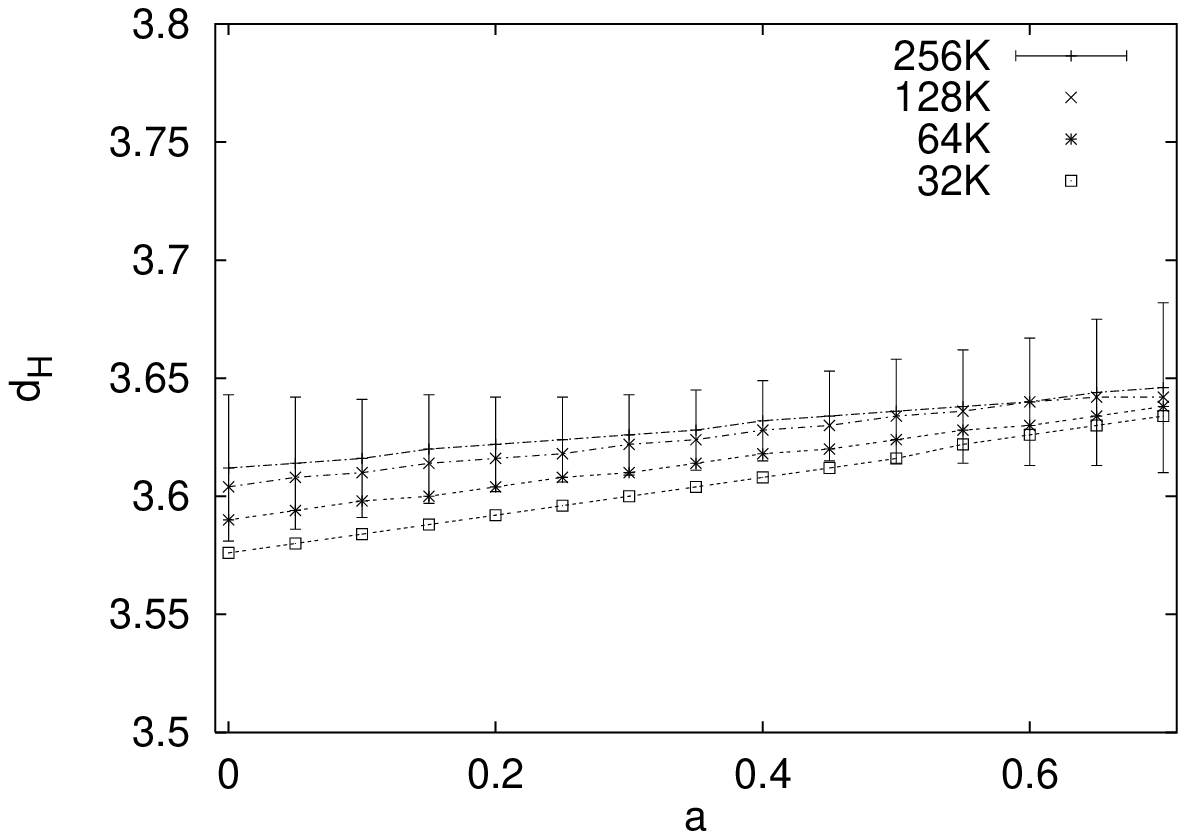}}
\centerline{\epsfxsize=4.0in\epsfysize=2.67in\epsfbox{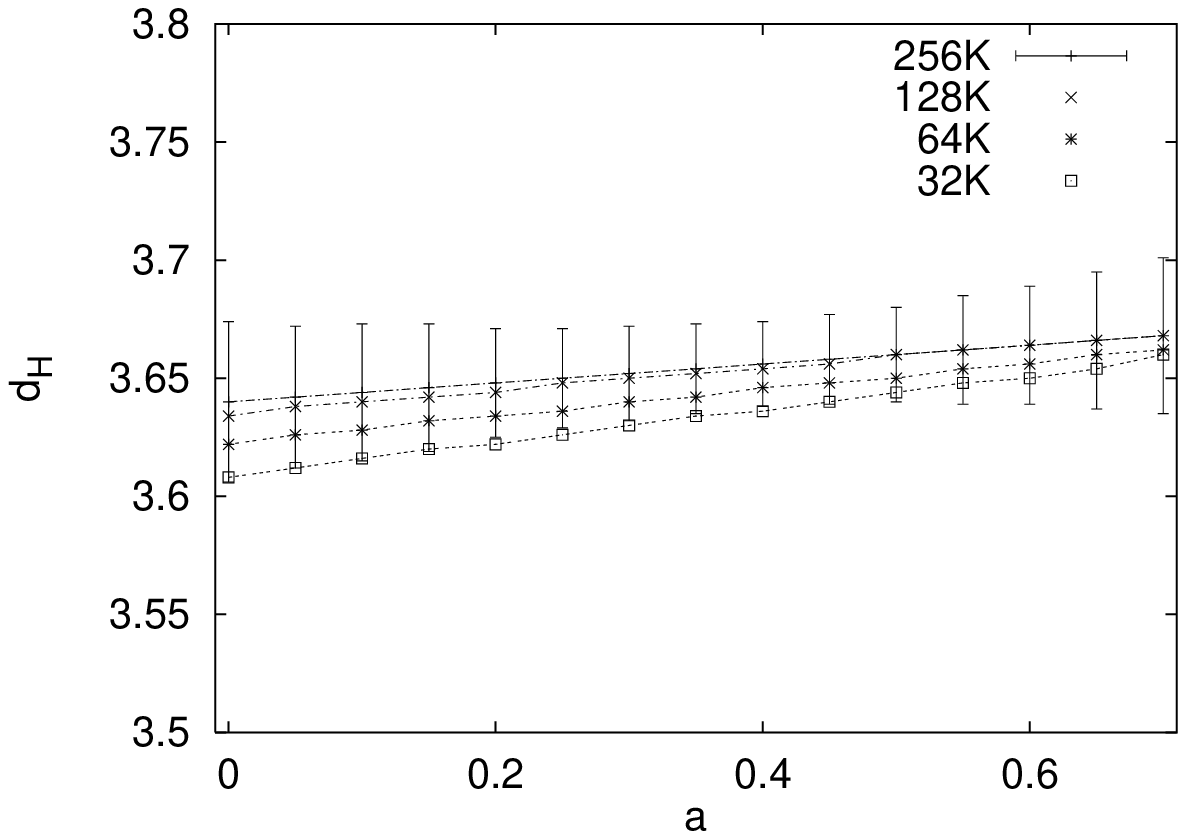}}
\centerline{\epsfxsize=4.0in\epsfysize=2.67in\epsfbox{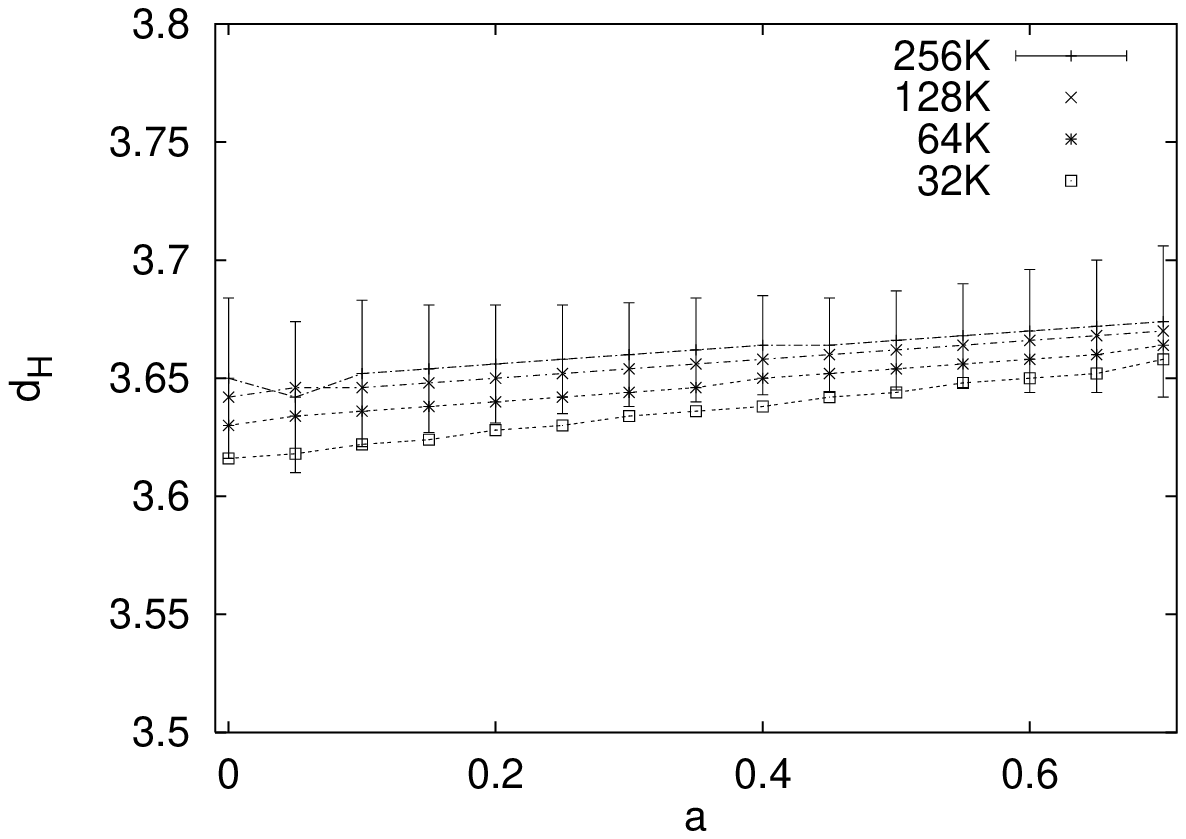}}
\caption{({\it a}) $d_H(a)$ computed from collapsing the $\vev{l^2}_{r,N}$
distributions. We collapse the distributions in groups of three,
indicating the largest lattice on the graph. 
({\it b}) The same as in ({\it a}) for $\vev{l^3}_{r,N}$. 
({\it c}) The same as in ({\it a}) for $\vev{l^4}_{r,N}$. }
\label{f:7}
\end{figure}

Let us now turn to the finite size scaling analysis of \rf{*21a}.
Again we introduce as a first phenomenological correction the shift $r
\to r+a$ as in \rf{*15a} to find the best scaling function $F_n(x)$
for a suitable range of $N_i$'s.  We have shown $F_n(x)$ for $n=2,3$
and 4 for the values of $a$ which provide the best scaling function in
fig.\ \ref{f:6}. The more detailed analysis is given in table
\ref{t:4}. Note that the results are perfectly consistent with the
same analysis for the first moment (table \ref{t:3}), but the shift
$a$ is different and in fact not very well determined (see
fig.~\ref{f:7}). In principle this is a good thing, but it implies
that we cannot use the shift in the same constructive way as for the
first moment and get a high precision measurement of $d_H$.

\begin{figure}[htb]
\centerline{\epsfxsize=4.0in \epsfysize=2.67in 
\epsfbox{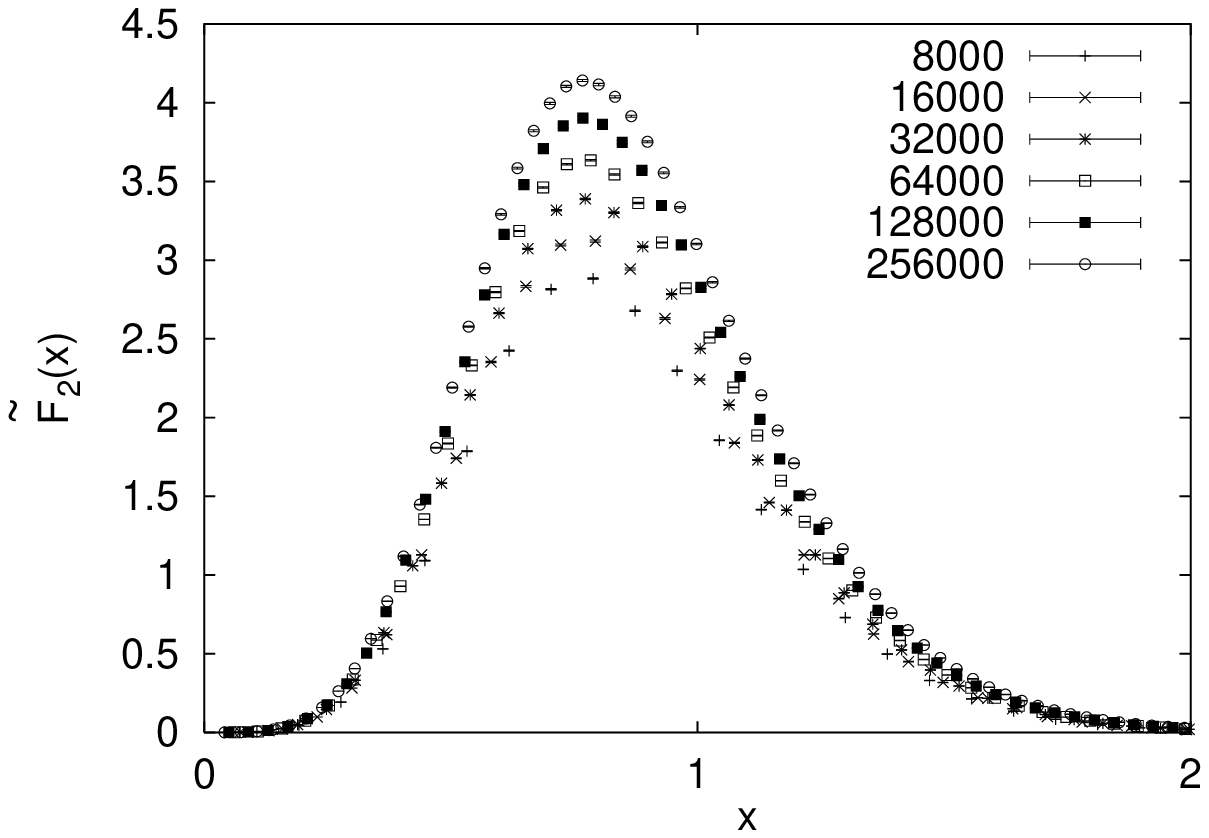}}
\centerline{\epsfxsize=4.0in \epsfysize=2.67in 
\epsfbox{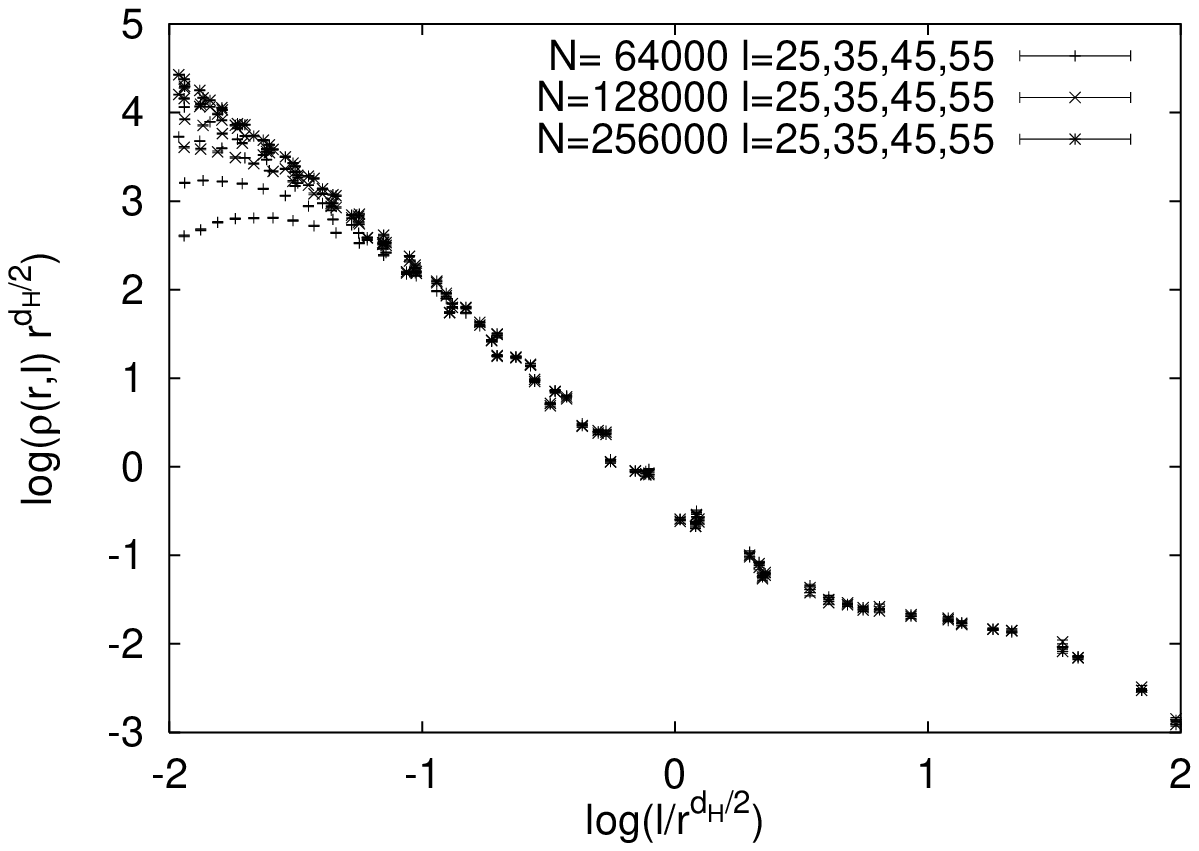}}
\caption{ The consequences of scaling using the assumption ${\rm
dim}[N]={\rm dim}[l^2]$: ({\it a}) The rescaled according to
Eq.~\protect\rf{*20a} $\vev{l^2}_{r,N}$. ({\it b}) The loop length
distribution function using $y=l/r^{d_H/2}$, $d_H=3.58$.}
\label{f:10}
\end{figure}

The scaling ansatz \rf{*20a} seems to be ruled out. In fig.\ \ref{f:10}a
we have shown the best overlap functions $\tilde{F}_2(x)$ 
for the second moment $\la l^2 \ra_{r,N}$ using  \rf{*20a}.
It should be compared to fig.\ \ref{f:6}, where we have used 
the ansatz \rf{*21a}. Clearly \rf{*21a} is superior to \rf{*20a}.
One may, however,  consider the possibility of a dimensional relation 
of the form: 
\beq{*22b} 
{\rm dim}[ \, l^n \, ]={\rm dim}[ \, r^{2n(1-\epsilon)} \, ] \, .
\eeq
In this case, for given $d_H$, one can use the relation 
$\la l^n \ra_{r,N} = N^{2n(1-\epsilon)/d_H}F_n(x)$ 
in order to determine the value of $\epsilon$.  
We get an upper bound on $\ep$ by using the lowest  value of 
for $d_H$ (obtained by some of the other methods discussed above)
and then fitting to the data. In this way we 
obtain $\epsilon < 0.03$. We consider the existence 
of such a small $\epsilon \neq 0$ unnatural.

\subsection{Distribution function}

Let us finally turn to the measurement of the distribution function
$\rho_N(r,l)$, which  provides the complete information
about the moments $\la l^n \ra_{r,N}$. 
In the case of pure gravity one can calculate $\rho_N(r,l)$ in the 
limit $N \to \infty$ and one finds \cite{transfer}:
\beq{*100}
\rho^{(c=0)}_\infty(r,l) \sim \frac{1}{r^2} G^{(c=0)}(l/r^2),
\eeq
where the function $G^{(0)}(z)$ is 
\beq{*101}
G^{(0)}(z) = \Bigl(\frac{1}{z^{5/2}}+\frac{1}{2 z^{3/2}} + 
\frac{14 z^{1/2}}{3} \Bigr)\; \e^{-z}.
\eeq
Note that this form of $\rho^{(0)}_\infty(r,l)$ explains
\rf{*90}. If $\ep$ denotes the cut-off (the lattice spacing 
in the triangulation) we can write
\beq{*102}
\la l^n \ra_{r,N=\infty}^{c=0} = 
\int_\ep  \d l \, l^n \rho^{(0)}_\infty(r,l) =
r^{2n} \int_{\ep /r^2} \d z\, z^n G^{(0)}(z).
\eeq
For $n > 1$ the lower limit has no implication for the integral and 
can be dropped. However, for $n=1$ the leading contribution in the 
limit $\ep \to 0$ comes precisely from the lower integration limit 
and we obtain:
\beq{*103}
\la l \ra_{r, N=\infty}^{c=0}  \sim \frac{1}{\ep^{1/2}}\; r^3.
\eeq
The cut--off dependence ensures 
that the real dimension of $\la l \ra$ is equal to that 
of $\sqrt{N}$.

Since we have verified with good accuracy that the higher moments for
$c=-2$ and $c=0$ has the {\em same} $r$ dependence for large $N$ (see
\rf{*90} and \rf{*22a}), we know that
\beq{*104}
\rho_N(r,l) \sim \frac{1}{r^2} G(l/r^2)
~~~~~
\hbox{for ~ $N \to \infty$}.
\eeq 
The function $G(z)$ need not be identical to $G^{(0)}(z)$.  In fact it
{\em cannot} be identical since $d_h(c=-2) \neq d_h(c=0)$.  Since we
expect that the origin of the different behaviour of the first and the
higher moments is the same for $c=0$ and $c=-2$ we know that the small
$z$ behaviour of $G(z)$ must be
\beq{*105}
G(z) \sim \frac{1}{z^{(d_h+1)/2}},
~~~~d_h \approx 3.57.
\eeq 

\begin{figure}[htb]
\centerline{\epsfxsize=4.0in\epsfysize=2.67in
\epsfbox{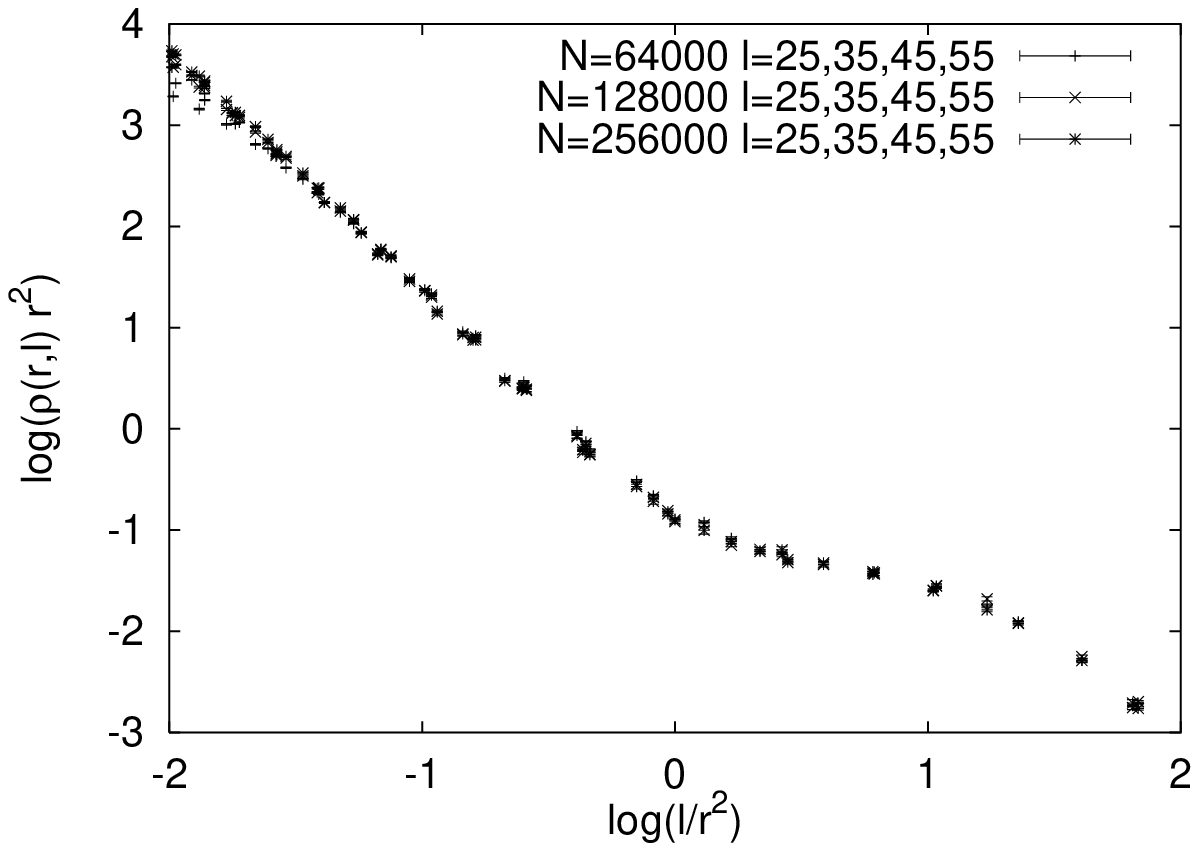}}
\centerline{\epsfxsize=4.0in\epsfysize=2.67in
\epsfbox{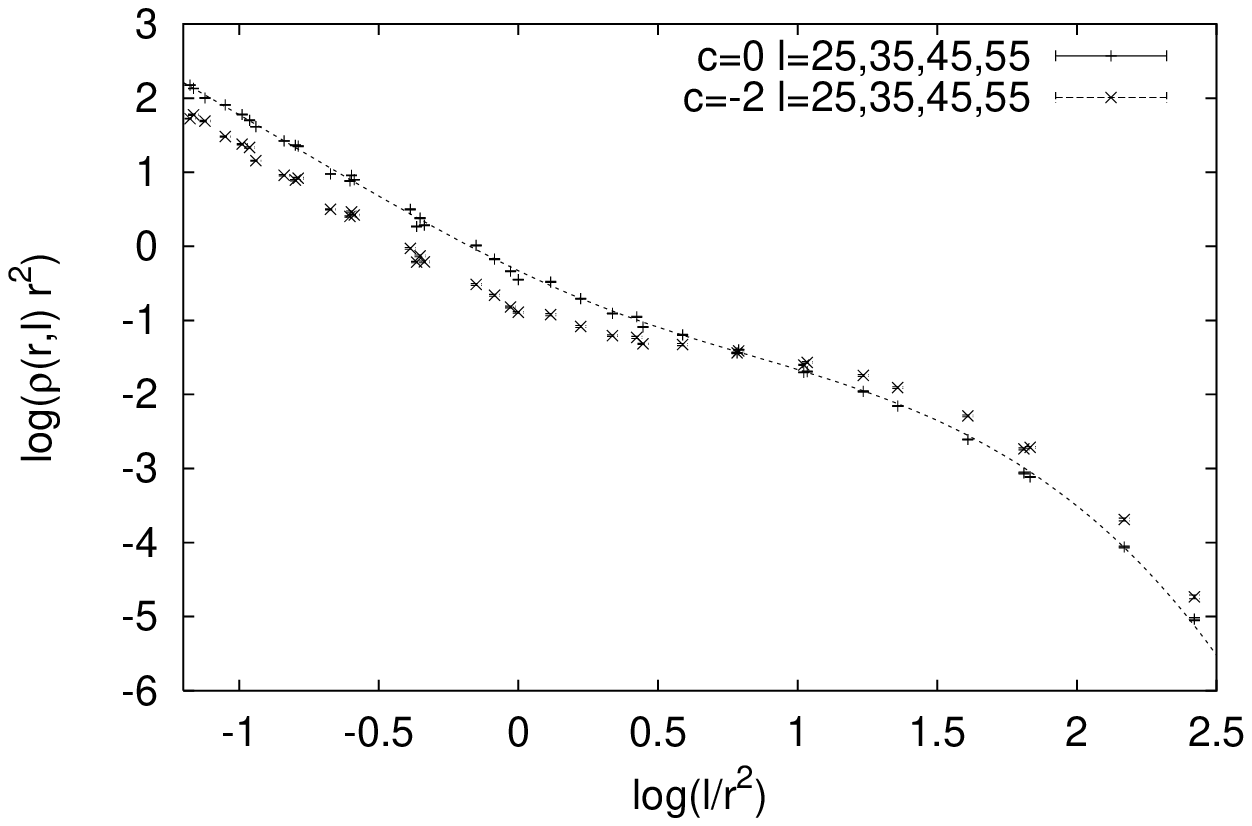}}
\centerline{\epsfxsize=4.0in\epsfysize=2.67in
\epsfbox{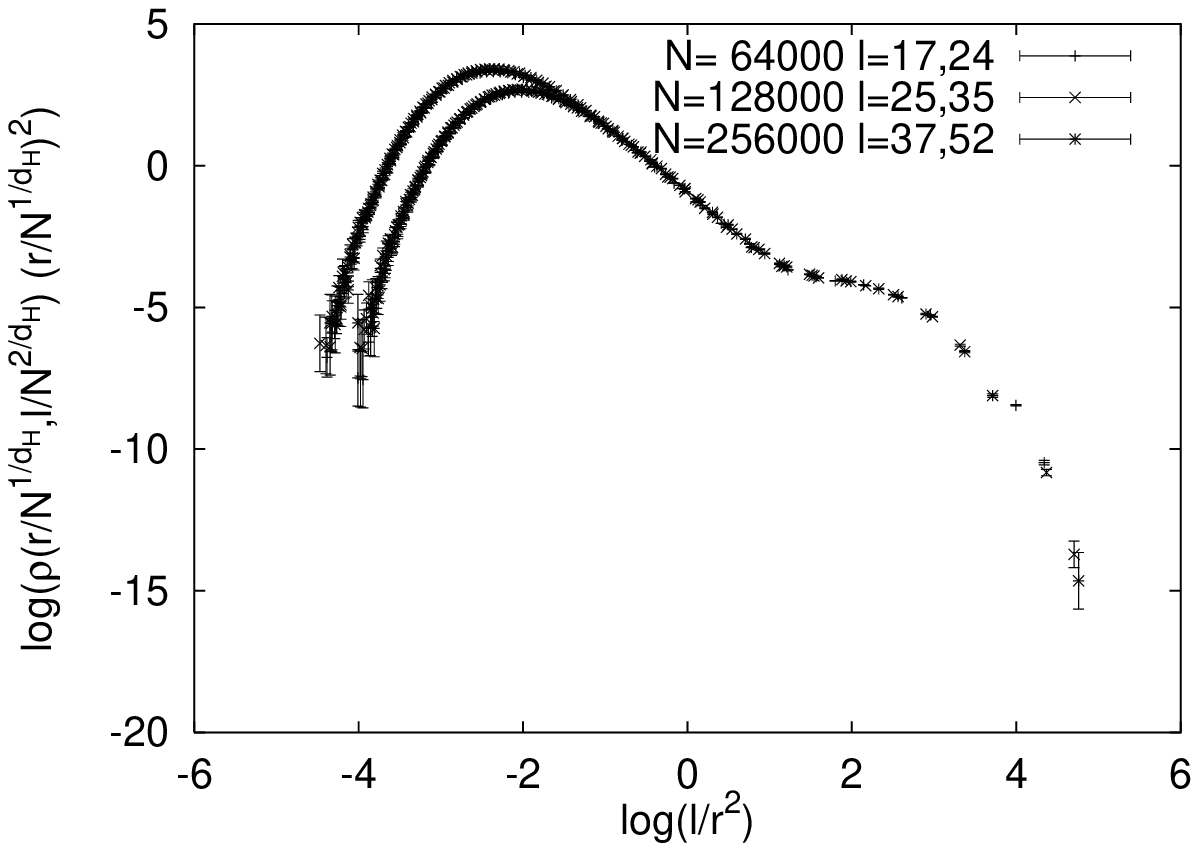}}
\caption{({\it a}) The loop length distribution $\rho_N(r,l)$ for
$N=64000,128000$ and $256000$.  
({\it b}) The same as in ({\it a}) for
$N=64000$ compared to the corresponding observable for pure
gravity ($c=0$). The dashed line is a fit to Eq.~\protect\rf{*101}.
The data for $c=0$ was taken from \protect\cite{ak}.
({\it c}) The same as in ({\it a}), but for (approximately) fixed 
$l/N^{2/d_H}$ and $d_H=3.58$.}
\label{f:9}
\end{figure}

Fig.\ \ref{f:9}a displays that the link-distance distribution 
function $r^2 \rho_N(r,l)$ is to a good 
approximation only a function  of $z=l/r^2$ as long as 
$r/N^{1/d_H} < 1$. For small values of $z$ the  $\log (G(z))$--curve 
is approximately a linear function of $\log (z)$. 
From \rf{*105} we expect
\beq{*106}
\log G(z) \sim 
- \, \frac{d_h+1}{2} \log z
~~~~~
\hbox{for ~ $\log z \to -\infty$}.
\eeq
A measurement of the slope gives $d_h = 3.60\pm 0.03$ in good 
agreement with short distance behaviour $\la l \ra_{r,N}$, as 
given in table \ref{t:3} for the link distance.
We emphasise that this agreement is no surprise. There has 
to be agreement. However, as we have seen, some treatments of the 
data set can produce very precise measurements of $d_h$. The slope 
of $\log G(z)$ does not belong to this class.

\begin{figure}[htb]
\centerline{\epsfxsize=4.0in\epsfysize=2.67in\epsfbox{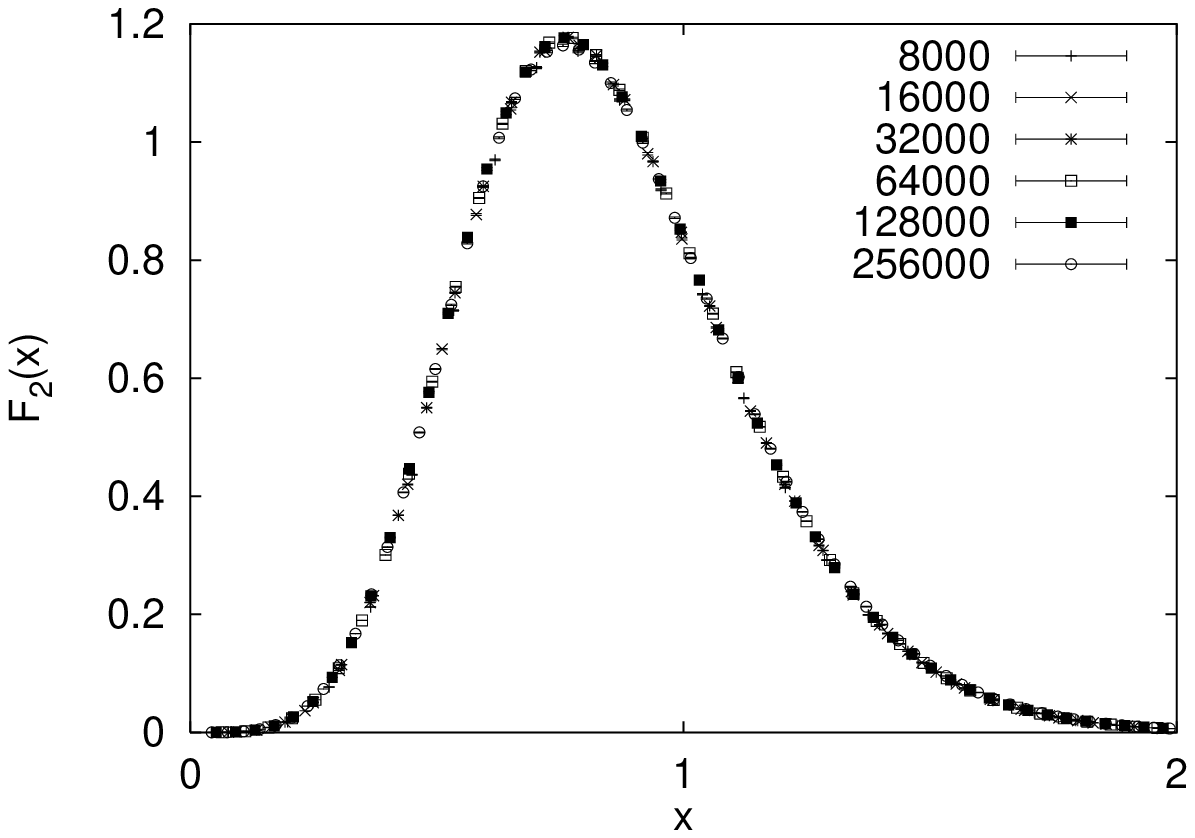}}
\centerline{\epsfxsize=4.0in\epsfysize=2.67in\epsfbox{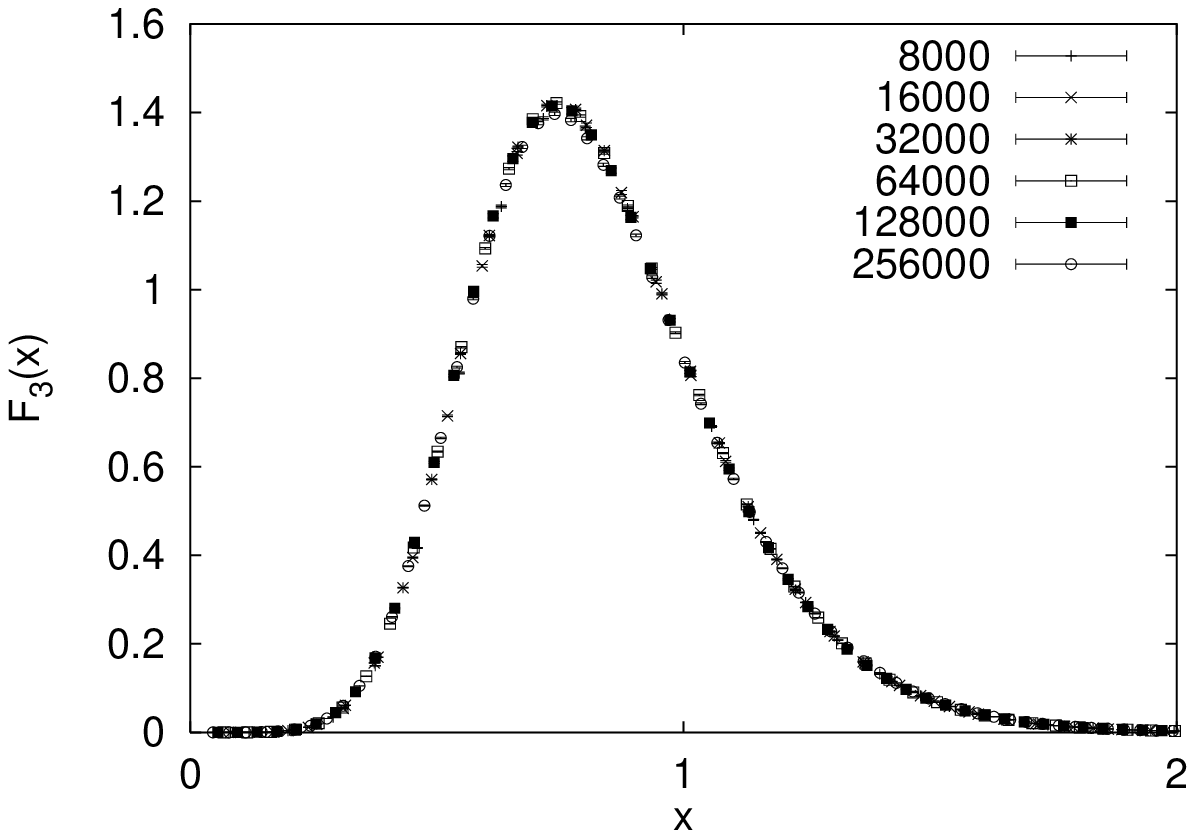}}
\centerline{\epsfxsize=4.0in\epsfysize=2.67in\epsfbox{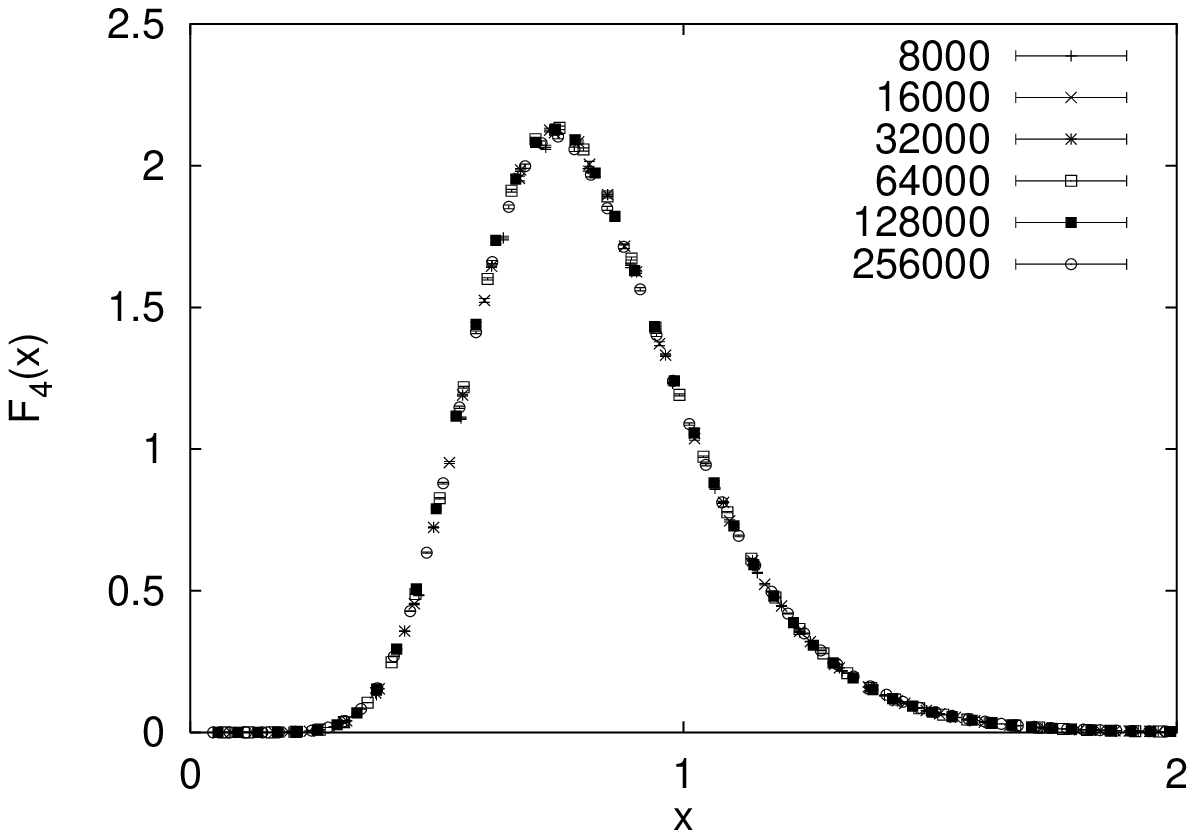}}
\caption{({\it a}) The $\vev{l^2}_{r,N}$ distributions defined in
terms of link distance rescaled according to Eq.~\protect\rf{*20a}
using $d_H=3.63$ and $a=0.35$.  ({\it b}) The same as in ({\it a}) for
$\vev{l^3}_{r,N}$, $d_H=3.65$ and $a=0.4$.  ({\it c}) The same as in
({\it a}) for $\vev{l^4}_{r,N}$, $d_H=3.66$ and $a=0.4$.}
\label{f:6}
\end{figure}

Finally, in fig.\ \ref{f:9}c, we have shown $r^2 \rho_N(r,l)$ in 
a somewhat larger $z$ interval. We know from measurements of the 
large $r$ behaviour of the moments $\la l \ra_{r,N}$ that 
they fall off fast when $x = r/N^{1/d_H} > 1$ (see \rf{*20a}
and fig.\ \ref{f:6}a-\ref{f:6}c). The same behaviour 
has to  be coded in $\rho_N(r,l)$. The simplest guess:
\beq{*107}
\rho_N(r,l) \sim \frac{1}{r^2} G(z) H(x)
\eeq
is not satisfied since this would imply that the scaling functions
$F_n(x)$ were proportional to $x^{2n} H(x)$, which is not very well
satisfied numerically.  However, the ansatz \rf{*107} is not too far
from the truth either. Since the curves shown in fig.\ \ref{f:9}c are
very similar to the ones for pure gravity one can test that an ansatz
like \rf{*107}, replacing $G(z)$ with $G^{(0)}(z)$ and $H(x)$ with the
function $f(x)$ from \rf{*72}.  The ansatz will produce curves
qualitatively very similar to the ones shown in fig.\ \ref{f:9}c.
Indeed, the curves consists of three major parts. Let $y=\log l/r^2$.
The right part of the curve goes as $-e^y$ and comes from the
exponential decay of $G^{(0)}(l/r^2)$ for positive $y$. The
``straight-line part'' is $-(5/2) y$ which dominates for negative $y$,
as discussed above. The ``bump'' is the joining of these two
asymptotic regions. Finally, for fixed $l/N^{2/d_H}$ and large
negative $y$ we hit the region where $f(x)$ goes to zero. It will
contribute with a term (for $ d_H=4$)
\beq{*108}
-\Bigl( \frac{l}{N^{1/2}} \Bigr)^{2/3}\e^{-\frac{2}{3} y}.
\eeq
This term depends on $l/N^{2/d_H}$
and consequently the decay of $\rho_N(r,l)$ for large negative 
$y$ will depend on $l/N^{2/d_H}$  precisely as is shown for 
the $c=-2$ case in fig.\ \ref{f:9}c.

\section{Theoretical estimate on the finite size shift 
of geodesic distance }
\label{shift}

In eq.~\rf{*81} we introduced the shift as part of a general finite
size expansion. Since it plays an important role in the fits it is
worthwhile to provide a theoretical understanding of the magnitude of
the shift. We shall limit ourselves to an analysis of the $c=0$ model
where it is possible to perform a theoretical analysis at a
discretized level.  There are other technical differences between the
theoretical and the numerical analyses which will be explained in
order.  Thus the comparison will be qualitative. Yet we believe that
the present analysis provides a theoretical confirmation of the
existence of a finite size shift of the geodesic distance.

The finite size shift $a$ will be shown to 
incorporate the effect of next higher-order correction 
in a lattice spacing parameter. 
Here, we denote the two-point functions 
at the discrete level and at the continuous level as 
$G_g(r)$ and $G_\mu(R)$, respectively. 
The two-point functions are related with 
$S_V(R)$ and $n_N(r)$ as 
\beq{GandS}
G_\mu(R) = \int\!dV e^{-\mu V} V Z(V) S_V(R) , 
~~~~~
G_g(r) = \sum_N g^{N} N Z_N \, n_N(r) , 
\eeq
where we have introduced 
$g$ as a coupling constant of matrix model, 
$r$ as the geodesic distance in the discrete level, 
$\mu$ as the cosmological constant, and 
$R$ as the geodesic distance at the continuous level. 
$g$ and $r$ are related with $\mu$ and $R$ by 
$g = g_c e^{- \alpha \ep^2 \mu}$ and $R = \beta r \ep^{2\nu}$, 
where $\alpha$ and $\beta$ are constant, and 
$\ep$ is the lattice spacing parameter in the triangulation. 

According to the arguments in ref.\ \cite{aw}, 
the continuum limit of the two-point function is 
\beq{G1}
G_g(r) = \ep^{2(\eta-1)\nu} \Bigl\{ 
G_\mu(R) + \ep^{2\nu} G_\mu^{(1)}(R) + O(\ep^{4\nu}) 
\Bigr\} \, , 
\eeq
where $\nu$ and $\eta$ are constant. 
{\it Suppose that the following relation 
\beq{katei1}
G_\mu^{(1)}(R)    %\propto %= \beta a %is proportional to 
= \beta a \frac{\partial}{\partial R} G_\mu(R) \, , 
\eeq
is satisfied,} we obtain 
\beq{G2}
G_g(r) = \ep^{2(\eta-1)\nu} \Bigl\{ 
G_\mu(R + \beta a \ep^{2\nu}) + 
O(\ep^{4\nu}) \Bigr\} \, ,
\eeq
where this $a$ will be identified 
as the finite size shift of geodesic distance. 
Thus, one can incorporate next higher-order correction 
by redefining the geodesic distance at the continuous level by 
$R^{\rm modified} = \beta (r+a) \ep^{2\nu}$ 
instead of by simply taking 
$R = \beta r \ep^{2\nu}$. 

Now, let us carry out a concrete calculation. 
Here, we restrict to the analysis of 
pure gravity ($c=0$ model) 
because this is the only case where 
the two-point function is theoretically known 
as a function of the geodesic distance $R$. 
In defining geodesic distance at the discrete level, 
there are two types of decomposition of triangles; 
slicing decomposition \cite{transfer} 
and peeling decomposition \cite{wata}. 
$G_\mu(R)$ is known for both cases 
while the discrete two-point function $G_g(r)$ 
is known only for the case of 
the peeling decomposition. 
In the peeling process, 
a triangle can be peeled off with 
$1/l$ step forward of geodesic distance 
($l$: loop length at the discrete level) 
while full one step forward of geodesic distance 
is realized in the one slicing process. 

In the peeling decomposition\cite{wata}, 
the two-point function %at the discrete level 
is related with the generating function of 
the disk amplitude at the discrete level \cite{aw}, 
%\footnote{
%This is the only example of two-point function described 
%at the discrete level.} 
\beq{TwoPointFun}
G_g(r) = 
\frac{\partial}{\partial x} F_g(\hat{x}(x,r)) \bigg|_{x=0} = 
\frac{1}{g} \frac{\partial}{\partial r} F_g(\hat{x}(0,r)) \, ,
\eeq
where we start from the boundary of disk with length $l_1=1$ at $r=0$.
Here, $F_g(x)=\sum_l x^l F_g(l)$ is the generating function of the
disk amplitude $F_g(l)$ with one marked link and boundary length $l$
at the discrete level.  The function $\hat{x}(x,r)$ satisfies
$\hat{x}(x,r=0) = x$ and
\beq{hatxdiffeq}
\frac{\partial \hat{x}}{\partial x} = 
\frac{f_g(\hat{x})}{f_g(x)} \, ,
~~~~~
\frac{\partial \hat{x}}{\partial r} = 
g f_g(\hat{x}) \, .
\eeq
In the case of one-matrix model\cite{BIPZ}, 
$F_g(x)$ and $f_g(x)$ have the following forms, 
\beq{Ftwofolded}
F_g(x) = 
\frac{1}{2} \biggl( \frac{1}{x^2} - \frac{g}{x^3} \biggr) + 
\frac{g}{2x^3} f_g(x) 
\, ,
~~~
f_g(x) = ( 1 - c_2 x ) \sqrt{ ( 1 - c_1 x ) ( 1 - c_0 x ) }
\, ,
\eeq
where $c_0 < 0 < c_1 < c_2$. 
The solution of \rf{hatxdiffeq} is\cite{aw} 
\beq{hatxtwofolded}
\hat{x}(x,r) = 
\frac{1}{c_2} \Biggl\{ 
1 - \frac{\delta_1}{ 
          \sinh^2\bigl( \delta_0 r + 
            \sinh^{-1}\!\sqrt{\frac{\delta_1}{1-c_2 x} - \delta_2} \bigr) 
          + \delta_2 } 
\Biggr\} \, ,
\eeq
where 
\beq{deltas}
\delta_0 = \frac{g}{2} \sqrt{ (c_2-c_1)(c_2-c_0) } , ~~~
\delta_1 = \frac{ (c_2-c_1)(c_2-c_0) }{ c_2(c_1-c_0) } , ~~~
\delta_2 = - \, \frac{ c_0(c_2-c_1) }{ c_2(c_1-c_0) } .
\eeq
Next, let us consider to take the continuum limit of \rf{TwoPointFun}. 
The continuum limit of the disk amplitude is 
taken by imposing 
$x = x_c e^{-\ep\zeta}$ and $g=g_c e^{- \alpha \ep^2 \mu}$, 
\bea
&&F_g(x) = {\rm (const.)} \ep^{3/2} \Bigl\{ 
- b_1 \ep^{-3/2} - b_2 \ep^{-1/2} \zeta + f_\mu(\zeta) + O(\ep^{1/2}) 
\Bigr\} \, , 
\label{Fcon}\\
&&~~~~~~
f_\mu(\zeta) = 
\Bigl( \zeta-\frac{\sqrt{\mu}}{2} \Bigr) \sqrt{\zeta+\sqrt{\mu}} \, ,
%~~~ \hbox{(for $x = x_c e^{-\ep\zeta}$ and $g=g_c e^{- \alpha \ep^2 \mu}$)}
\eea
where the constants, $b_1$ and $b_2$, and 
the critical values, $x_c$ and $g_c$, are regularization dependent. 
The continuum limit of $\hat{x}(0,r)$ is 
\beq{hatxcon}
\hat{x}(0,r) = 
x_c \Bigl\{ 1 - \ep \hat{\zeta}_0(R + \beta a_1 \sqrt{\ep}) 
            + O(\ep^2) \Bigr\} \, ,
\eeq
where 
\beq{hatzeta}
\hat{\zeta}_0(R) = 
- \sqrt{\mu} + \frac{3}{2} \sqrt{\mu} 
\coth^2 \biggl( \sqrt{\frac{3}{2}}\mu^{1/4} R \biggr) \, , 
\eeq
and $R = \beta r \sqrt{\ep}$, 
which means that $\nu=1/4$ in pure gravity. 
Substituting \rf{Fcon} and \rf{hatxcon} into \rf{TwoPointFun}, 
we find 
\beq{G3}
G_g(r) = {\rm (const.)} \ep^{3/2} 
\frac{\partial}{\partial R} \biggl\{ 
   - b_2 \hat{\zeta}_0(R + \beta a_1 \sqrt{\ep}) + 
   \sqrt{\ep} f_\mu\bigl(\hat{\zeta}_0(R)\bigr) + O(\ep) 
\biggr\} \, , 
\eeq
which tells us that $\eta=4$. 
Here, we obtain the following nontrivial relation from \rf{hatzeta}, 
\beq{Dzf}
\frac{\partial}{\partial R} \hat{\zeta}_0(R) = 
 -2 f_\mu(\hat{\zeta}_0(R)) \, .
\eeq
Then we find the property \rf{katei1}, i.e., 
we finally find 
\beq{Gfinal}
G_g(r) = 
- {\rm (const.)} \ep^{3/2} \frac{\partial}{\partial R} \biggl\{ 
\hat{\zeta}_0\Bigl(R + \beta a_1 \sqrt{\ep} + \frac{\sqrt{\ep}}{2b_2}\Bigr) 
+ O(\ep) 
\biggr\} \, . 
\eeq
Thus, we obtain 
\beq{Rmodified1}
R^{\rm modified} = \beta (r+a) \ep^{2\nu} 
~~~\hbox{where}~~~ 
a = a_1 + a_2, 
\biggl( a_2 = \frac{1}{2b_2\beta} \biggr) \, .
\eeq
In the one-matrix model which describes pure gravity, 
the critical values are 
$x_c = (3^{1/4}-3^{-1/4})/2$ and $g_c = 1/(2\cdot 3^{3/4})$. 
Then, we find 
$b_2 = \sqrt{3}(\sqrt{3}+1)^{-3/2}$, 
$\beta = (\sqrt{3}+1)^{1/2}/(2\sqrt{3})$, and 
$a_1 = \sqrt{3}$.
Therefore, the total finite size shift becomes 
\beq{Shift1}
a = a_1 + a_2 = 2 \sqrt{3} + 1 = 4.46...
~~~
( a_1=\sqrt{3},~ a_2=\sqrt{3}+1 ) \, .
\eeq

We now try to evaluate the two-point function where the starting
boundary length is $3$ at $r=0$.  In this case we have to take the
third derivative of the generating function of the disk amplitude
instead of the first derivative of \rf{TwoPointFun},
%from $\partial/\partial x$ into $(\partial/\partial x)^3$. 
\bea
G^{[l_1=3]}_g(r) 
&=& \frac{1}{3!} 
    \biggl( \frac{\partial}{\partial x} \biggr)^3 F_g(\hat{x}) \bigg|_{x=0} 
\label{TwoPointFun3}\\
&=& \frac{1}{3!} \Biggl\{ 
    \biggl( \frac{1}{f_g} \biggr)'' 
  + \frac{3}{g} \biggl( \frac{1}{f_g} \biggr)' 
    \frac{\partial}{\partial r} 
  + \frac{1}{g^2} 
    \biggl( \frac{\partial}{\partial r} \biggr)^2 
    \Biggr\} 
    \frac{1}{g} \frac{\partial}{\partial r} F_g(\hat{x}) \bigg|_{x=0} \, .
\nonumber
\eea
The continuum limit of \rf{TwoPointFun3} is 
\beq{G4}
G_g^{[l_1=3]}(r) 
=  {\rm (const.)} \biggl( 1 + 
    \beta a_3 \sqrt{\ep} \frac{\partial}{\partial R} + O(\ep) \biggr) 
     G_g^{[l_1=1]}(r) 
\, , 
\eeq
where 
$G_g^{[l_1=1]}(r)$ is $G_g(r)$ of \rf{Gfinal} and 
\beq{a3}
a_3 \ = \  \lim_{\ep \rightarrow 0} 
           \frac{3( 1/f_g )'}{g( 1/f_g )''} \bigg|_{x=0} \, .
\eeq
In the one-matrix model, the value of $a_3$ is $a_3 = 3/2$. 
For the case with the boundary disk length $l_1=3$ at $r=0$, 
we obtain the total finite size shift, 
\beq{Shift2}
a = a_1 + a_2 + a_3 = 2 \sqrt{3} + \frac{5}{2} = 5.96... \, .
\eeq

The similar calculation can be done for the triangulation without
two-folded links (two links are on top of each other) which come from
the kinetic term in the matrix model.  In this case the disk amplitude
is
\beq{Fnotwofolded}
F_g(x) = 
\frac{x}{2g} - \frac{1}{2} - x^2 + \frac{1}{2} f_g(x)
\, ,
~~~
f_g(x) = ( 1 - c_2 x ) \sqrt{ 1 - c_1 x }
\, ,
%~~~
%c_1 = \frac{4}{g c_2^2} , 
%~
%\frac{c_2}{g^2} - c_2^3 = \frac{8}{g} , 
\eeq
where $c_1 = 4/(g c_2^2)$ and $c_2^3 - c_2/g^2  = -8/g$. 
Since $f_g(x)$ in \rf{Fnotwofolded} is formally the same as 
$f_g(x)$ in \rf{Ftwofolded} with the substitution $c_0=0$, 
$\hat{x}(x,r)$ of the present model is formally the same as 
$\hat{x}(x,r)$ in \rf{hatxtwofolded} with $c_0=0$. 
%The function $\hat{x}(x,r)$ is 
%\beq{xhat2}
%\hat{x}(x,r) = 
%\frac{1}{c_1} \Biggl\{ 
%1 - \frac{\delta_1}{ 
%          \tanh^2\bigl( \delta_0 r + 
%            \tanh^{-1}\!\sqrt{\frac{\delta_1}{1-c_1 x}} \bigr) } 
%\Biggr\} \, ,
%\eeq
%where 
%\beq{delta2}
%\delta_0 = \frac{1}{2} \sqrt{ 1 - \frac{12g}{c_2} } , ~~~
%\delta_1 = 1 - \frac{c_1}{c_2} .
%\eeq
In this case, the continuum limit is realized around the critical
values, $x_c = 1/(2\cdot 3^{1/4})$ and $g_c = 1/(2\cdot 3^{3/4})$.
The continuum limit of $F_g(x)$ is \rf{Fcon} with $b_2 = 2/\sqrt{3}$,
and that of $\hat{x}(0,r)$ is \rf{hatxcon} with $\beta =
1/(2\sqrt{3})$ and $a_1 = 2\sqrt{3}$.  Then, for the present model we
find
\beq{Shift3}
a = a_1 + a_2 = 2 \sqrt{3} + \frac{3}{2} = 4.96...
~~~
( a_1=2\sqrt{3},~ a_2=\frac{3}{2} ) \, .
\eeq
In the present case with the boundary disk length $l_1 = 3$ at $r=0$, 
the total finite size shift is 
\beq{Shift4}
a = a_1 + a_2 + a_3 = \frac{16\sqrt{3}}{5} + \frac{3}{2} = 7.04... 
\, ,
\eeq
where $a_3 = 6\sqrt{3}/5$ which is derived from \rf{a3}. 

In the slicing decomposition in pure gravity\cite{transfer}, we have
$a_2 = \sqrt{(38+21\sqrt{3})/6}$, because of the property that
\rf{Dzf} is independent of regularization.  However, we failed to
obtain $a_1$ because we do not know the concrete expression of
$\hat{x}(x,r)$ in the slicing decomposition.

\begin{figure}[htb]
\begin{center}
\subfigure{
\psfig{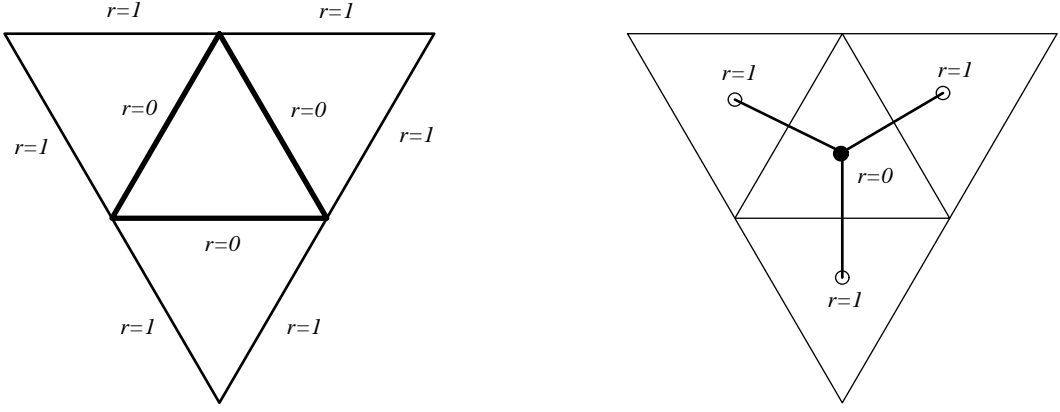} }
\end{center}
\caption{({\it a}) The distance used in the theoretical analysis. 
         ({\it b}) The distance used in our numerical simulation.
         The difference of the distances 
         between ({\it a}) and ({\it b}) is about $1/2$.}
\label{distance}
\end{figure}

The above arguments show that the finite size shift $a$ depends on the
detailed prescription for the starting point and the class of
triangulations used. As a further source of ambiguity we can mention
that the counting of distance in the theoretical considerations above
and in numerical simulations differ in the following way: In our
numerical simulations we first mark one of the triangles on the sphere,
and then measure the distance from the marked triangle.  The marked
triangle might be a normal triangle or a triangle which corresponds to
a tadpole in the dual lattice.  Thus the marked point is located at
the center of the marked triangle in our simulation while the boundary
of the marked triangle is the zero geodesic region in the above
theoretical analysis.
%Removing the starting triangle from the sphere, 
%one gets a disk with the boundary length 
%$l_1=1$     %(if the marked triangle is tadpole)  
%or $l_1=3$. %(otherwise)  
%On the other hand, 
%the above geodesic distance $r$ is measured 
%from the boundary of a disk with length $l_1=1$ or $l_1=3$. 
In fig.\ \ref{distance} we describe the definitions of the distance
both in the above theoretical analysis and in our numerical analyses.
Therefore, the geodesic distance in the numerical simulation $r^{\rm
sim}$ is related with the geodesic distance $r$ as $r^{\rm sim} \sim
r+1/2$.  Since $R^{\rm modified} = \beta (r+a) \ep^{2\nu} = \beta
(r^{\rm sim}+a^{\rm sim}) \ep^{2\nu}$, which reproduces the universal
function $G_\mu(R^{\rm modified})$, we find $a^{\rm sim} \sim a-1/2$.
%Even more important is the fact that in the theoretical approach we
%use the peeling decomposition for the definition of geodesic distance
%whereas in the numerical simulation we use the slicing decomposition.
%Here, it should be noted that the geodesic distance of the present
%theoretical analysis is defined as the shortest path on the dual
%lattice while the numerical analyses include both of the definitions;
%the shortest path on the dual lattice and on the original lattice.
%So, we should compare the above theoretical results with the numerical
%results measured by using the dual lattice geodesic distance.  
%There is an important difference between the theoretical and
%the numerical analyses.  In the theoretical analysis peeling
%decomposition is used to estimate the geodesic distance $r$ since this
%is the only case where the discrete two-point function is known.  In
%the numerical simulation the slicing technique is used to measure the
%geodesic distance $r^{\rm sim}$.  Therefore this may cause some
%quantitative difference, i.e.\ the theoretically predicted $a^{\rm
%sim}_{\rm th}$ and numerical $a^{\rm sim}_{\rm num}$ could be
%quantitatively different.
Another slight difference between the theoretical and numerical
determination of the shift $a$ is the different definition used for
the boundary lengths. In the numerical simulations we count the number
of triangles at a given distance $r^{\rm sim}$ whereas in the
theoretical approach we count the number of links at distance
$r$. Finally, in the computer simulations we use a modified version of
the slicing decomposition. This implies that we cannot directly
compare with the theoretical results derived above which use the
peeling decomposition to estimate the geodesic distance $r$. However,
when we compare with the numerical simulations with $c=0$ it is found
that the shift is $4<a<6$ \cite{ajw} which is consistent with
eqs.~\rf{Shift1},\rf{Shift2},\rf{Shift3} and \rf{Shift4}.

%To summarise, however, the theoretically predicted finite size shift
%$a^{\rm sim}$ for the dual lattice geodesic distance is found to be
%more or less {\it about\/} $4 \sim 6$, depending on the choice of
%models and initial distances, which is consistent with the previous
%result of numerical simulations for the $c=0$ model \cite{ajw}.

For more general models where $c\neq 0$, we have not succeeded in
showing the property \rf{katei1} which is used in proving the
existence of the shift $a$.  On the other hand, the numerical
simulation in $c\neq 0$ models strongly supports the existence of the
shift $a$ of the same order of magnitude as for the $c=0$ model.

Finally, we consider the shift in $\langle l^n \rangle_r$. 
In the peeling decomposition, $\langle l^n \rangle_r$ is described by 
\beq{Ln}
\langle l^n \rangle_r = 
\frac{\partial}{\partial x} 
\biggl( \hat{x} \frac{\partial}{\partial \hat{x}} \biggr)^{n-1} 
F_g(\hat{x}(x,r)) \bigg|_{x=0}  \, ,
\eeq
where we start from the boundary of disk with length $l_1=1$ at $r=0$. 
Note that from \rf{hatxcon} we obtain 
\beq{hatxcon2}
\hat{x} \frac{\partial}{\partial \hat{x}} 
\ = \ 
- \, \frac{1}{\ep} \frac{\partial}{\partial \hat{\zeta}_0} + O(\ep) \, .
\eeq
Substituting \rf{hatxcon2} into \rf{Ln} 
and carrying out the similar calculation as before, 
we find for $n \ge 2$, 
\beq{Ln2}
\langle l^n \rangle_r = {\rm (const.)} \ep^{3-n} 
\frac{\partial}{\partial R} 
\biggl( \frac{\partial}{\partial \hat{\zeta}_0} \biggr)^{n-1} 
\biggl\{ 
  f_\mu\bigl(\hat{\zeta}_0(R + \beta a_1 \sqrt{\ep})\bigr) 
  + ({\rm const.}) \sqrt{\ep} \Bigl( \hat{\zeta}_0(R) \Bigr)^2 
  + O(\ep) 
\biggr\} \, . 
\eeq
For $n \ge 3$ the finite size shift comes only from $a_1$ because
$a_2$ is absent.  When we start from the boundary of disk with length
$l_1=3$ at $r=0$, the shift is $a_1 + a_3$. The values of $a$ for
$\vev{l^n}_r$ ($n\ge 3$) are all equal but are smaller than that of
$a$ for $\vev{l}_r$ by $a_2$.  On the other hand, for $n=2$ the
property \rf{katei1} is broken, because $\hat{\zeta}_0^2$ is not
proportional to $(\partial/\partial R) f_\mu(\hat{\zeta}_0)$.  So, we
cannot expect a clear shifting property for $\langle l^2 \rangle_r$ in
the numerical simulations.

\section{Discussion}

In this article we have taken advantage of the special recursive sampling 
possible for the $d=-2$ theory coupled to quantum gravity. The quality of 
the numerical data obtained this way is much better than the 
quality of data 
obtained with a similar computer effort by ordinary Monte Carlo simulations. 
The purpose of the present work has been to measure the fractal structure of 
space-time in the $c=-2$ theory coupled to quantum gravity with a precision 
which has not been available before, by combining the high quality 
data with the technique of finite size scaling.      

In this way we obtained, with a conservative error estimate,
\begin{equation}\label{*disc1}
d_H = 3.58\pm 0.04,~~~~~d_h = 3.56\pm 0.04,
\end{equation}
in perfect agreement with the theoretical prediction $3.561\cdots$ 
from the diffusion in Liouville theory (see (\ref{*25}) and (\ref{*29})), 
and in disagreement with the prediction given by (\ref{*1})-(\ref{*3}).

Thus eq.\ (\ref{*25}) is strongly favoured as the correct formula for
the fractal dimension of space-time in the case where the matter
fields coupled to gravity have $c < 0$. In this region eq.\
(\ref{*25}) has nicer properties than eq.\ (\ref{*1}), since one
naively expects that $d_H(c) \to 2$ for $c \to -\infty$.  It is
reassuring that (\ref{*25}), rather than (\ref{*1}), is selected as
the correct formula, since $d_H$ in (\ref{*25}) goes to 2 for $d \to
-\infty$. In addition (\ref{*disc1}) provides strong evidence for the
existence of a unique fractal dimension $d_H = d_h$ at all distances.
As shown in the Appendix one cannot take for granted such a relation,
in particular for non--unitary theories coupled to quantum gravity,
but based on (\ref{*disc1}) it is tempting to conjecture that,
contrary to the situation for the so-called multi-critical branched
polymers discussed in the Appendix, we will always have $d_H = d_h$ in
quantum gravity.

We have emphasised the validity of (\ref{*25}) for $c \leq 0$, but
have remained silent about the region $0 < c < 1$.  The reason is that
for non--unitary conformal field theories we have operators with
negative scaling dimensions. They will be dominant relative to the
cosmological term, and it cannot be entirely ruled out that a formula
like (\ref{*1}) is correct for $c >0$.  Contrary to eq.\ (\ref{*25})
it has a drastic dependence on $c$ for $c \to 1$. This dependence has
not been observed until now in the computer simulations which favour
$d_H \approx 4$.  For $c >0$ the numerical data are based on ordinary
Monte Carlo simulations and it is not possible to perform the same
high precision determination of $d_H$ as for $c=-2$. The best present
data are not in good agreement with (\ref{*25}), but cannot falsify it
either.

Eqs.\ (\ref{*21a}) and (\ref{*22a}) show a new and surprising scaling
indicating that $l/r^2$ seems to be a universal, dimensionless
variable. In the case $c=0$ this is certainly reasonable since we know
that the dimension of $r$ is $N^{1/4}$ and it is natural to expect
that a boundary has dimension $N^{1/2}$. This is {\it not} satisfied
for $\langle l \rangle_r$ due to the fractal structure of space-time,
which implies that the boundary of a sphere of geodesic radius $r$ is
highly multi-connected with many microscopic loops. This manifests
itself in the short distance cut-off needed in eq.\ (\ref{*102}) for
$n=1$.  However, for moments $\langle l^n \rangle_r$, $n >1$, these
microscopic loops play no role and we get $\langle l^n \rangle_r \sim
N^{n/2}$ (see eq.\ (\ref{*91})). If $l/r^2$ is a universal,
dimensionless variable also for $c < 0$ where $d_H < 4$, we reach the
conclusion that the $l$ appearing in the computer measurements {\em
cannot} be identified with the ``Liouville'' $\ell$, which in the
continuum notation is believed to have the dressing dictated by:
\begin{equation}
\delta(\int\! d^2\xi \sqrt{\hat{g}}\; e^{\alpha \phi}- V) \to 
\delta( \int\! d \xi \sqrt[4]{\hat{g}}\; e^{\alpha \phi /2} -\ell ),
\label{*disc2}
\end{equation}
where $\alpha$ denotes the gravitational dressing associated 
with the cosmological term and the gravitational dressing of 
the boundary cosmological term is $\alpha/2$.
In the context of Liouville theory it has never been entirely clear 
how to derive this result from first principles since the actual 
boundary conditions of the matter fields for $c < 0$ have never 
been explicitly specified. In the numerical simulations the boundary 
is not fixed, but only characterised by being a (multi-connected) spherical 
shell of geodesic radius $r$. Probably {\em free} boundary conditions in 
Liouville theory come closest to the ``experimental'' set-up in the 
numerical simulations, and maybe the existence of scaling operators with 
negative dimensions in the non-unitary theories can spoil the 
relationship between the dimension of space-time and boundary shown 
in eq.\ (\ref{*disc2}) in the case of free boundary conditions, 
since these operators are dominant relative to the cosmological constant.   

We hope it will eventually  be possible to understand this new, 
observed  scaling of the boundary length  from first principles. 
   
\subsection*{Acknowledgements}
J. Ambj\o rn acknowledges the support of the Professor Visitante Iberdrola 
Grant and the hospitality at the University of Barcelona, where part
of this work was done.
Y. Watabiki acknowledges the support 
and the hospitality of the Niels Bohr Institute. 
J. Ambj\o rn and N. Kawamoto were supported by the Exchange program of 
Japanese Ministry of Education, Science and Culture, under the 
Grand-in-Aid number 07044048.  

\section*{Appendix}

The purpose of this appendix is to provide an example of a statistical 
ensemble to which one can assign both kind of fractal dimensions,
$d_H$ and $d_h$ as discussed in sec.\ 2.3, but where they do not coincide.
The model we have in mind is a model of so-called multi-critical 
branched polymers \cite{adj}. It is defined as a  statistical ensemble of 
connected, planar  graphs without any loops, i.e.\ connected planar 
tree-graphs.  For such a tree-graph the weight is given by 
a fugacity factor $e^{-\mu}$ for each link $l$ and a branching factor 
$f(n_v)$ for each vertex $v$ of order $n_v$. If $BP$ denotes 
the class of planar tree graphs,  the partition function 
is given by 
\begin{equation}
Z^{(2)}_\mu = \sum_{G\in BP} \frac{1}{{\cal S}_G} 
\prod_{v\in G} f(n_v) \prod_{l \in G} e^{-\mu}, 
\label{*app1}
\end{equation}
where ${\cal S}_G$ is a symmetry factor for the graph $G$, 
such that a  {\em rooted} branched polymer is counted only once.
It is known that for a large class of positive functions $f$ the 
first derivative of the partition function $Z^{(2)}_\mu$ is given by 
\begin{equation}
Z^{(2)}_\mu{}' = c_2 - (\mu -\mu_2)^{1/2}~~~~~{\rm for}~~~\mu \to \mu_2,
\label{*app2}
\end{equation} 
where $\mu_2$ and $c_2$ are non-universal constants.
The geodesic distance between two vertices in a tree graph is defined 
as the shortest link distance between the two vertices. The two-point 
function, defined as in eq.\ (\ref{*app1}), except that two marked 
points are separated a geodesic distance $r$, can be calculated and is 
given by 
\begin{equation}
G^{(2)}_\mu(r) = \Bigl[Z^{(2)}_\mu{}'\Bigr]^2 \; 
e^{- k_2 \,r \sqrt{\mu-\mu_2}}, 
\label{*app3}
\end{equation} 
for $\mu \to \mu_2$
where $k_2$ is again a non-universal constant.  
{}From (\ref{*app2}) and (\ref{*app3}) 
we can find the partition function as well as 
the two-point function for constant volume,
i.e.\ for a constant number of links, by a (discrete) Laplace transformation
in $\mu$: 
\begin{equation}
Z^{(2)}_N \sim N^{-5/2}\times e^{\mu_2 N},
\end{equation}
\begin{equation}
G^{(2)}_N (r) \sim N^{-3/2}\; r \, e^{-\tilde{k}_2 \, r^2/N}\times e^{\mu_2 N}.
\label{*app4}
\end{equation}
The ``spherical shell'' of geodesic radius $r$ for graphs of volume $N$,
$n^{(2)}_N(r)$,  is 
defined by analogue with (\ref{*70}) and counts the average number of 
vertices of distance $r$ from an arbitrarily chosen vertex:
\begin{equation}
n^{(2)}_N(r) = 
\frac{ G^{(2)}_N(r)}{N Z^{(2)}_N} \sim r\, e^{-\tilde{k}_2 \, r^2/N}.
\label{*app5}
\end{equation}
Comparing (\ref{*70})-(\ref{*74}) with (\ref{*app5}) we conclude that 
$d_H = d_h =2$ for this class of branched polymers.

There exists a generalisation of the generic class of branched polymers 
considered here, which in many respects are related  to 
the generic branched polymers as the multi-critical matrix models
are related to pure two-di\-men\-si\-onal gravity. They are called 
{\em multi-critical branched polymers} and as for multi-critical 
matrix models they are defined by allowing certain negative weights for
some of the orders $n_v$ of vertices. By fine-tuning of the weights one 
can obtain a new critical behaviour:
\begin{equation}
Z^{(m)}_\mu{}' = c_m - (\mu -\mu_m)^{1/m}~~~~~{\rm for}~~~\mu \to \mu_m,
\label{*app6}
\end{equation}
\begin{equation}
G^{(m)}_\mu(r) = \Bigl[Z^{(m)}_\mu{}'\Bigr]^2 \; e^{- k_m\, r (\mu-\mu_m)^{1-1/m}}, 
\label{*app7}
\end{equation} 
where $c_m$, $\mu_m$ and $k_m$ are non-universal constants and 
$m > 2$ ($m\!=\! 2$ corresponds to the generic branched polymer). 
{}From eq.\ (\ref{*app7}) it follows that $d_H = m/(m-1)$. 
However,  inverse Laplace
transformations lead to the following  expressions (in the limit of 
large $N$) for the partition function with fixed volume $N$,
\begin{equation}
Z^{(m)}_N  \sim  N^{-2-1/m}\times e^{\mu_m N}, 
\end{equation}
and the two-point function with fixed volume $N$,
\begin{eqnarray}
G^{(m)}_N (r) &= & \int_{-i\infty}^{i\infty} 
d\mu \; e^{\mu N} G^{(m)}_\mu (r) \nonumber\\
&=&\int_{-i\infty}^{i\infty} d\mu \; e^{\mu N}
\Bigl( c_m \!-\! (\mu\!-\!\mu_m)^{1/m}\Bigr)^2 
\Bigl( 1\! -\! k_m\, r (\mu\!-\! \mu_m)^{1-1/m} \!+\! \cdots\Bigr) \nonumber\\
&=& N Z^{(m)}_N \Bigl( \tilde{c}_0 + \tilde{c}_1\frac{r}{N^{1-2/m}} + 
O(r^2) \Bigr).
\label{*app7a}  
\end{eqnarray}
{}From this equation we conclude that
\begin{equation}
n^{(m)}_N(r) \sim \frac{r}{N^{1-\frac{2}{m}}}~~~~{\rm for}~~~
N^{1-2/m} \ll r \ll N^{1-1/m}.
\label{*app8}
\end{equation}  
This shows that from a formal point of view 
$d_h = 2$, and  one has the situation 
indicated in eq.\ (\ref{*71a}), with $d_h=2$ and $d_H = m/(m-1)$.

\newpage 

%\begin{table}[ht]
%\begin{center}
%\begin{tabular}{}
%\hline

%\hline
%\end{tabular}
%\end{center}
%\caption{}
%\label{t:1}
%\end{table}

\begin{table}[ht]
\begin{center}
\begin{tabular}{| r |  l l l  |c c c|}\hline\multicolumn{1}{|c |}{$N$}& 
     \multicolumn{1}{c }{$d_h$} & 
     \multicolumn{1}{c }{$a$}   & 
     \multicolumn{1}{c|}{$C$}
      & $\chi^2$& $r_{\rm min}$& $r_{\rm max}$ \\
\hline
8192000& 3.5335(7) & 4.277(8) & 0.0429(1) & 1.2 &  5 & 105\\
2048000& 3.5253(6) & 4.222(6) & 0.0443(1) & 1.1 &  5 & 60\\
1024000& 3.5166(8) & 4.184(6) & 0.0456(2) & 1.5 &  5 & 45\\
 512000& 3.498(2)  & 4.08(2)  & 0.0489(4) & 1.0 &  5 & 45\\
 256000& 3.4932(6) & 4.061(4) & 0.0497(1) & 1.7 &  5 & 30\\
 128000& 3.483(3)  & 4.020(8) & 0.0515(3) & 1.2 &  5 & 15\\
  64000& 3.464(1)  & 3.945(6) & 0.0546(2) & 0.5 &  5 & 15\\
8192000& 3.533(1)  & 4.27(2)  & 0.0430(2) & 1.2 & 10 & 105\\
2048000& 3.525(1)  & 4.21(1)  & 0.0445(2) & 1.1 & 10 & 60\\
1024000& 3.529(1)  & 4.22(1)  & 0.0452(3) & 1.1 & 10 & 50\\
 512000& 3.490(4)  & 3.99(4)  & 0.0507(8) & 0.9 & 10 & 45\\
 256000& 3.490(1)  & 4.03(1)  & 0.0504(3) & 1.4 & 10 & 30\\
 128000& 3.463(2)  & 3.88(1)  & 0.0555(3) & 3.6 & 10 & 25\\
\hline
\end{tabular}
\end{center}
\caption{The fractal dimension $d_h$ as determined from the small
distance scaling of $n_N(r)$ when $r$ is the triangle distance.}
\label{t:2}
\end{table}

\begin{table}[ht]
\begin{center}
\begin{tabular}{| r |  l l l  |c c c|}\hline\multicolumn{1}{|c |}{$N$}& 
     \multicolumn{1}{c }{$d_h$} & 
     \multicolumn{1}{c }{$a$}   & 
     \multicolumn{1}{c|}{$C$}
      & $\chi^2$& $r_{\rm min}$& $r_{\rm max}$ \\
\hline
512000& 3.593(5) & 0.480(9) & 1.72(2) & 1.0& 2& 10\\
256000& 3.625(3) & 0.530(5) & 1.59(1) & 1.5& 2&  6\\
128000& 3.623(3) & 0.530(4) & 1.59(1) & 1.7& 2&  5\\
 64000& 3.607(2) & 0.512(3) & 1.645(7)& 8.7& 2&  5\\
512000& 3.556(6) & 0.40(1)  & 1.91(3) & 1.2& 3& 12\\
256000& 3.594(5) & 0.476(9) & 1.72(2) & 0.8& 3&  7\\
128000& 3.584(4) & 0.465(8) & 1.75(2) & 1.0& 3&  6\\
 64000& 3.544(3) & 0.408(5) & 1.92(1) & 6.9& 3&  6\\
512000& 3.525(7) & 0.31(2)  & 2.10(5) & 1.0& 4& 13\\
256000& 3.560(7) & 0.40(2)  & 1.89(4) & 0.7& 4&  8\\
128000& 3.535(6) & 0.36(1)  & 2.00(3) & 1.0& 4&  7\\
\hline
\end{tabular}
\end{center}
\caption{The fractal dimension $d_h$ as determined from the small
distance scaling of $n_N(r)$ when $r$ is the link distance.}
\label{t:3}
\end{table}

\begin{table}[ht]
\begin{center}
\begin{tabular}{|l l |l l | r c l |}
\hline 
\multicolumn{2}{|c }{link dist.} &
\multicolumn{2}{|c }{triangle dist.} &
\multicolumn{3}{|c|}{}\\
\cline{1-4}
\multicolumn{1}{|c }{$d_H$} & 
\multicolumn{1}{ c }{$a$} & 
\multicolumn{1}{|c }{$d_H$} & 
\multicolumn{1}{ c }{$a$} &  
\multicolumn{3}{|c|}{\raisebox{1.5ex}[0pt]{$N$}}\\
\hline
3.602(20) & 0.50(15) & 3.560(16) & 4.60(35) & 256000 &{--}& 128000 \\
3.610(14) & 0.50(7)  & 3.552(10) & 4.50(20) & 128000 &{--}&  64000 \\
3.621(10) & 0.54(5)  & 3.538(6)  & 4.30(10) &  64000 &{--}&  32000 \\
3.634(4)  & 0.55(1)  & 3.520(8)  & 4.20(10) &  32000 &{--}&  16000 \\
\hline
3.608(12) & 0.50(7)  & 3.555(5)  & 4.55(15) & 256000 &{--}&  64000\\
3.612(7)  & 0.50(4)  & 3.544(4)  & 4.40(10) & 128000 &{--}&  32000\\
3.630(8)  & 0.55(3)  & 3.532(8)  & 4.30(15) &  64000 &{--}&  16000\\
\hline
3.610(8)  & 0.50(5)  & 3.549(7)  & 4.45(15) & 256000 &{--}&  32000\\
3.618(5)  & 0.52(5)  & 3.538(12) & 4.35(20) & 128000 &{--}&  16000\\
\hline
3.575(8)  & 0.30(5)  & 3.573(8)  & 5.0(2)   & 256000 &{--}&  ${\rm 16000}^*$\\
\hline
\end{tabular}
\end{center}
\caption{The fractal dimension $d_H$ as determined from collapsing 
$n_N(r)$. Both definitions of distance are included. ${ }^*$ refers
to the graphical computation of $d_H$ from the intersection points of
the $d_H(a)$ curves shown in Fig.~\protect\ref{f:3}}
\label{t:1}
\end{table}

\begin{table}[ht]
\begin{center}
\begin{tabular}{|l l |l l | r c l |}
\hline 
\multicolumn{2}{|c }{link dist.} &
\multicolumn{2}{|c }{triangle dist.} &
\multicolumn{3}{|c|}{}\\
\cline{1-4}
\multicolumn{1}{|c }{$d_H$} & 
\multicolumn{1}{ c }{$a$} & 
\multicolumn{1}{|c }{$d_H$} & 
\multicolumn{1}{ c }{$a$} &  
\multicolumn{3}{|c|}{\raisebox{1.5ex}[0pt]{$N$}}\\
\hline
 3.571(2) & 0.131(4) & 3.570(5) & 4.93(7) & 256000 &{--}&  2000\\
 3.573(12)& 0.138(30)& 3.574(20)& 5.02(40)& 256000 &{--}&  4000\\
 3.574(3) & 0.137(7) & 3.573(3) & 4.97(6) & 256000 &{--}&  8000\\
 3.575(4) & 0.141(7) & 3.576(3) & 5.04(7) & 256000 &{--}& 16000\\
 3.578(7) & 0.150(25)& 3.574(7) & 5.00(17)& 256000 &{--}& 32000\\
\hline
\end{tabular}
\end{center}
\caption{The fractal dimension $d_H$ as determined from collapsing 
the intersection point of $R_{a,N}(d)$ for $c=-2$. 
Both definitions of distance are included.}
\label{t:6}
\end{table}

\begin{table}[ht]
\begin{center}
\begin{tabular}{|c | r |  l l l  |c c c|}\hline
     $n$ &
     \multicolumn{1}{|c |}{$N$}& 
     \multicolumn{1}{c }{$2 n$} & 
     \multicolumn{1}{c }{$a$}   & 
     \multicolumn{1}{c|}{$C$}
      & $\chi^2$& $r_{\rm min}$& $r_{\rm max}$ \\
\hline 
2 & 512000 & 4.12(1)  & 0.310(8) & 11.6(3) & 1.2 & 1 & 5 \\
  & 256000 & 4.117(7) & 0.312(4) & 11.6(2) & 4.1 & 1 & 5 \\
  & 128000 & 4.10(1)  & 0.309(7) & 11.8(3) & 3.1 & 1 & 5 \\
  &  64000 & 4.122(8) & 0.317(4) & 11.4(2) & 4.4 & 1 & 4 \\
  & 512000 & 3.95(1)  & 0.17(1)  & 16.8(5) & 1.2 & 2 & 8 \\
  & 256000 & 3.98(1)  & 0.21(1)  & 15.5(4) & 2.8 & 2 & 6 \\
  & 128000 & 3.94(2)  & 0.18(2)  & 16.7(7) & 0.9 & 2 & 6 \\
  &  64000 & 3.93(1)  & 0.17(1)  & 17.2(5) & 2.6 & 2 & 5 \\
\hline
3 & 512000 & 6.09(2) & 0.326(9) & 0.65(3)$\times{\rm10}^{\rm2}$& 1.3 & 1 & 6 \\
  & 256000 & 6.14(1) & 0.345(5) & 0.59(2)$\times{\rm10}^{\rm2}$& 2.5 & 1 & 5 \\
  & 128000 & 6.11(2) & 0.342(9) & 0.61(3)$\times{\rm10}^{\rm2}$& 2.4 & 1 & 5 \\
  &  64000 & 6.14(2) & 0.348(6) & 0.58(2)$\times{\rm10}^{\rm2}$& 2.5 & 1 & 4 \\
  & 512000 & 5.91(2) & 0.21(2)  & 1.01(6)$\times{\rm10}^{\rm2}$& 0.9 & 2 & 8 \\
  & 256000 & 5.94(2) & 0.23(1)  & 0.93(5)$\times{\rm10}^{\rm2}$& 2.8 & 2 & 6 \\
  & 128000 & 5.86(4) & 0.20(2)  & 1.07(9)$\times{\rm10}^{\rm2}$& 0.6 & 2 & 6 \\
  &  64000 & 5.83(3) & 0.19(1)  & 1.12(6)$\times{\rm10}^{\rm2}$& 2.5 & 2 & 5 \\
\hline
4 & 512000 & 8.02(2) & 0.325(9) & 0.50(3)$\times{\rm10}^{\rm3}$& 2.1 & 1 & 8 \\
  & 256000 & 8.18(2) & 0.373(7) & 0.36(2)$\times{\rm10}^{\rm3}$& 1.4 & 1 & 5 \\
  & 128000 & 8.13(4) & 0.37(1)  & 0.37(1)$\times{\rm10}^{\rm3}$& 1.6 & 1 & 5 \\
  &  64000 & 8.01(2) & 0.336(6) & 0.47(2)$\times{\rm10}^{\rm3}$& 10  & 1 & 5 \\
  & 512000 & 7.81(3) & 0.22(2)  & 0.85(7)$\times{\rm10}^{\rm3}$& 1.3 & 2 & 9 \\
  & 256000 & 7.90(4) & 0.26(2)  & 0.68(6)$\times{\rm10}^{\rm3}$& 2.5 & 2 & 6 \\
  & 128000 & 7.64(4) & 0.16(2)  & 1.2(1) $\times{\rm10}^{\rm3}$& 1.9 & 2 & 7 \\
  &  64000 & 7.74(4) & 0.21(2)  & 0.90(9)$\times{\rm10}^{\rm3}$& 2.0 & 2 & 5 \\
\hline
\end{tabular}
\end{center}
\caption{The exponent $2n$ as determined from the small
distance scaling of $\vev{l^n}_{r,N}$.}
\label{t:5}
\end{table}

\begin{table}[ht]
\begin{center}
\begin{tabular}{| c  | l l | r c l |}
\hline
$n$ &
\multicolumn{1}{|c }{$d_H$} & 
\multicolumn{1}{ c }{$a$} &  
\multicolumn{3}{|c|}{$N$}\\
\hline
2 & 3.629 (33) & 0.35 (20) & 256000 &{--}& 64000\\
  & 3.616 (23) & 0.20 (10) & 128000 &{--}& 32000\\
  & 3.606 (18) & 0.25 (10) &  64000 &{--}& 16000\\
  & 3.588 (14) & 0.15 (5)  &  32000 &{--}&  8000\\
\hline
3 & 3.654 (28) & 0.40 (20) & 256000 &{--}& 64000\\
  & 3.645 (25) & 0.23 (17) & 128000 &{--}& 32000\\
  & 3.636 (21) & 0.27 (10) &  64000 &{--}& 16000\\
  & 3.621 (20) & 0.20 (10) &  32000 &{--}&  8000\\
\hline
4 & 3.662 (40) & 0.40 (30) & 256000 &{--}& 64000\\
  & 3.648 (32) & 0.20 (20) & 128000 &{--}& 32000\\
  & 3.641 (33) & 0.25 (15) &  64000 &{--}& 16000\\
  & 3.626 (30) & 0.20 (15) &  32000 &{--}&  8000\\
\hline
\end{tabular}
\end{center}
\caption{The fractal dimension $d_H$ as determined from collapsing 
$\vev{l^n}_{r,N}$.}
\label{t:4}
\end{table}

\clearpage

\newpage 

%\begin{figure}[htb]
%\centerline{\epsfxsize=4.0in \epsfysize=2.67in \epsfbox{}}
%\caption{}
%\label{f:1}
%\end{figure}

\clearpage

%\begin{figure}[htb]
%\begin{center}
%\subfigure{
%\psfig{figure=sphere.eps,width=8cm,angle=90} }
%\end{center}
%\vspace{24pt}
%\caption{Partition function described by tree and rainbow diagrams.}
%\label{tree0}
%\end{figure}

%\begin{figure}[htb]
%\begin{center}
%\subfigure{
%\psfig{figure=tree_a.eps,width=12cm,angle=90} }
%\end{center}
%\begin{center}
%\subfigure{
%\psfig{figure=tree_b.eps,width=14cm,angle=90} }
%\end{center}
%\caption{({\it a}) Identity for tree diagrams. 
%         ({\it b}) Identity for rainbow diagrams.}
%\label{tree}
%\end{figure}

%\begin{figure}[htb]
%\begin{center}
%\subfigure{
%\psfig{figure=distance.eps,width=14cm,angle=90} }
%\end{center}
%%\begin{center}
%%\subfigure{
%%\psfig{figure=distance_a.eps,width=6cm,angle=90} }
%%\subfigure{
%%\psfig{figure=distance_b.eps,width=6cm,angle=90} }
%%\end{center}
%\caption{({\it a}) The distance used in the theoretical analysis. 
%         ({\it b}) The distance used in our numerical simulation.
%         The difference of the distances 
%         between ({\it a}) and ({\it b}) is about $1/2$.}
%\label{distance}
%\end{figure}

\clearpage

\end{document}